\documentclass[a4paper,11pt]{article}
\pdfoutput=1 

\usepackage{jheppub} 

\usepackage[T1]{fontenc} 

\usepackage{amssymb}

\usepackage{epsfig,amsmath,amssymb
}

\newcommand{\be}{\begin{equation}}
\newcommand{\ee}{\end{equation}}

\newcounter{unnumber}

\newtheorem{thrm}{Theorem}
\newtheorem{lm}{Lemma}
\newtheorem{cor}{Corollary}
\newtheorem{prop}{Proposition}

\newtheorem{de}{Definition}
\newtheorem{prf}[unnumber]{Proof}

\font\tenscr=rsfs10 scaled1100
\font\sevenscr=rsfs7 
\font\fivescr=rsfs5 
\skewchar\tenscr='177 \skewchar\sevenscr='177 \skewchar\fivescr='177
\newfam\scrfam
\textfont\scrfam=\tenscr
\scriptfont\scrfam=\sevenscr
\scriptscriptfont\scrfam=\fivescr

\def\scri{{\fam\scrfam I}}

\title{\boldmath Representations of the
Bondi$-$Metzner$-$Sachs  group
in three space$-$time dimensions in the Hilbert topology
I. Determination of the representations}

\author[a
]{Evangelos Melas
}

\affiliation[a]{University of Thessaly,\\Department of Mathematics,\\Lamia 35100, \\ Greece}

\emailAdd{evanmelas@uth.gr}

\abstract{The  original  Bondi$-$Metzner$-$Sachs (BMS) group $B$ is the common asymptotic symmetry group of all asymptotically flat Lorentzian 4$-$dim space$-$times. As such, $B$ is  the best candidate for the universal symmetry group of General Relativity (G.R.).
Here,
the analogue $B(2,1)$  of $B$
in 3 space$-$time dimensions is properly defined.
We study its representation theory in the Hilbert topology
by using an infinite$-$dimensional extension of Wigner$-$Mackey theory.
We obtain the necessary data in order to construct the strongly continuous irreducible unitary representations  (IRS)
of $B(2,1).$
The main results of the representation theory are:
The IRS are induced from ``little groups'' which are compact.
There is one infinite connected ``little group'', the special orthogonal group
$SO(2).$ There are infinite non$-$connected finite discrete ``little groups'',
the cyclic groups $C_{n}$ of even order.
The inducing construction is exhaustive notwithstanding the fact that $B(2,1)$
is not locally compact in the employed Hilbert topology.
$B(2,1)$ is also derived, in Klein's sense, as the automorphism group of the
``strong conformal geometry'' of $\scri^{+}$.
Besides the Hilbert topology
other reasonable
topologies are given to $B(2,1)$ and their physical relevance is discussed.
Connections of the IRS of $B(2,1)$ with geometry are outlined.

\vspace{1cm}
}

\begin{document}
\maketitle
\flushbottom

\section{Introduction}

\label{intro}
The
BMS group $B$ is the common asymptotic group of all
curved real Lorentzian space$-$times which are  asymptotically flat in future null directions \cite{Bondi,Sachs1}, and is the best candidate for the universal symmetry group of G.R..

The BMS group $B$ of general relativity was first
discovered \cite{Bondi,Sachs1} not as a transformation group of
(exact) global diffeomorphisms of a fixed manifold,
but as a pseudo$-$group of local diffeomorphisms
(``asymptotic isometries") of the asymptotic region
of Lorentzian space$-$times which are asymptotically
fiat in lightlike future directions.

However, Penrose \cite{Pen1,Pen2,Pen3} showed that, by ``going to infinity", $B$ could be interpreted as an (exact) global transformation group $B \times \Im^{+} \longrightarrow \Im^{+}$ of the ``future null boundary" $\Im^{+}$ of the space$-$times concerned. Furthermore, he gave \cite{Pen1,Pen2}  a geometric structure to $\Im^{+}$, the ``strong conformal geometry" for which the
transformation group is precisely the group of automorphisms
\cite{Pen3}.

In 1939 Wigner laid the foundations
of special relativistic quantum mechanics \cite{Wigner} and relativistic quantum field theory by constructing
the Hilbert space
IRS of the (universal cover) of the Poincare group $P$.

The universal property of $B$ for G.R.
makes it reasonable to attempt
to lay a similarly firm foundation for quantum gravity by following through the analogue of
Wigner's programme with $B$ replacing $P$. Some years ago McCarthy constructed explicitly \cite{mac3,mac8,mac4,mac5,mac6,mac7,Crampin2,Mac}   the IRS of $B$ for exactly this purpose.
This work was based on G. W. Mackey's pioneering work on group representations \cite{Mackey0,Mackey00,Mackey,Mackey1,Simms,Isham}; in particular
it was based on an extension \cite{mac6,Mel6} of G. W. Mackey's  work to the relevant
infinite dimensional case.


It is difficult to overemphasize the importance of Piard's results \cite{Piard1,Piard2} who soon afterwards proved that \it{all} \normalfont the IRS of $B$,
when this is equipped with the Hilbert topology, are derivable by the inducing construction.
This proves the exhaustivity of McCarthy's list of representations and renders his results even more important.

However, in quantum gravity, complexified or euclidean versions of G.R. are frequently considered and
the question arises: Are there similar symmetry groups for these versions of the theory? McCarthy constructed \cite{mac1}, in abstract form,
all possible analogues of $B$, both real and in any signature, or complex, with all possible notions of asymptotic flatness ``near infinity''. There are, in fact, forty$-$one such groups.
One of them is $B(2,2)$, the asymptotic symmetry group of all curved real ultrahyperbolic space$-$times which are asymptotically flat in null directions.

These abstract constructions were given in a quantum setting; the paper \cite{mac1} was concerned with finding the IRS of these groups
in Hilbert spaces (especially for the complexification $ \mathcal C \mathcal B $ of $B$ itself). It has been argued \cite{mac1,Mel1}
that these Hilbert space representations are
related to elementary particles and quantum gravity (via gravitational instantons). The IRS of $B(2,2)$ were constructed in \cite{macMel,Mel5}.
The representation theory of HB, which is a subgroup of $B(2,2)$, was initiated in \cite{Melas1,Mel6a,Mel7}.

Here,
we follow this programme for 3$-$dim G.R.
We properly define the analogue $B(2,1)$ of $B$
in three space$-$time dimensions.
We construct the IRS
of $B(2,1)$
by endowing it
with the Hilbert topology.
IRS
in the Hilbert topology describe
\it bounded \normalfont sources \cite{Crampin2}.

$B(2,1)$ was first introduced in \cite{Mel6}.
However, the treatment of $B(2,1)$ in \cite{Mel6} was rather discursive.
Only the main features of $B(2,1)$ were given without any relevant proofs.
\cite{Mel6} dealt primarily with the needed
infinite$-$dimen$-$\newline
sional extension of Wigner$-$Mackey theory
required to study the representation theory $B$ and all its variants
in 3 space–time dimensions  and higher in all signatures.
In this paper, the  extension of Wigner$-$Mackey theory given in \cite{Mel6},
is  particularized  in the case of $B(2,1).$
Moreover, further particulars of the necessary extension
of Wigner$-$Mackey theory appropriate to the case of
$B(2,1)$ are given mainly regarding the various reasonable topologies
$B(2,1)$ can be equipped with and their physical underpinnings.

The main results of the representation theory  are:
The
IRS of
$B(2,1)$
are induced
from IRS of \it compact \normalfont ``little groups''.
The ``little groups''
are of two types:
1. Infinite connected Lie groups. 2. Non$-$connected finite discrete groups.
The infinite  connected ``little group'' is the special orthogonal group $SO(2).$
The  non$-$connected finite discrete ``little groups'' are
cyclic groups of even order, which are symmetry groups of  regular
polygons in ordinary euclidean 2$-$space.
The inducing construction is exhaustive
notwithstanding the fact that $B(2,1)$
is
\it not \normalfont
locally compact in the employed Hilbert topology.
We give all the necessary information
in order to construct the operators of the induced representations of
$B(2,1)$.


$B(2, 1)$ is the semidirect product of $SL(2, R)$ with the additive group of real$-$valued functions defined on
the circle. The asymptotic symmetry group in 3 space–time dimensions, to the best of our knowledge,
was first considered in \cite{Bi}, a work which is entirely
classical in aims and scope. Later in \cite{Ba1} the authors defined, in three space$–$time dimensions, $BMS_{3}$,
 the semidirect product of $Diff(S^{1})$, the group of the orientation
preserving diffeomorphisms of the circle, with the additive group of real$-$valued functions defined on
the circle. In \cite{Ba2}, \cite{Ba3} the projective representations of $BMS_{3}$ were studied.
 The main difference between this paper and \cite{Bi}, \cite{Ba1}, \cite{Ba2}, \cite{Ba3} is that in
 \cite{Bi}, \cite{Ba1}, \cite{Ba2}, \cite{Ba3}, $SL(2, R)$ has been enlarged to the full group   $Diff(S^{1})$.
As a result  there is no overlap between this paper and \cite{Bi}, \cite{Ba1},  \cite{Ba2}, \cite{Ba3}.


This paper is organised as follows:
In section \ref{Group} the group $B^{2,1}(N^{+})$ is introduced.
In section \ref{dc} the double cover $B^{2,1}(N^{+})_{c}$ of
$B^{2,1}(N^{+})$  is defined.
In section \ref{closeup} the group $B^{2,1}(N^{+})_{c}$ is studied.
In section  \ref{anB} the group  $B(2,1)$ is defined.
In section  \ref{auto}  the group  $B(2,1)$ is rederived as
automorphism group of the strong conformal geometry of $\scri^{+}$.
In section \ref{classical} the classical action of $B(2,1)$
on $\scri^{+}$ is given. In section \ref{subgroup} the subgroup
of supertranslations of $B(2,1)$   is studied.
In section \ref{subgroups} the subgroup structure of
$B(2,1)$ is analyzed.
In section \ref{spaces} the spaces $\mathcal A, \   \mathcal A^{\prime}, \  \hat{\mathcal A}  $
are defined.
In section \ref{dualaction} the dual action of the action which defines $B(2,1)$
on
$\mathcal A^{\prime}$ is given.
In section \ref{representation} Wigner$-$Mackey's theory is outlined and
its necessary extension is presented.
In section \ref{topologies}  a class of ``reasonable'' topologies
for $\mathcal A$  is given. In section \ref{compact} it is proved that all
``Little groups'' are compact. In section \ref{littlegroups} the ``Little groups'' of
$B(2,1)$ are given.
In section \ref{exhaustitivity} it is proved that
 \it all \normalfont IRS of $B(2,1)$ are obtained by the Wigner$-$Mackey's  inducing construction.
In section \ref{construction} all the necessary data in order to construct the IRS of $B(2,1)$ are given.
In section \ref{geometry}  connections of the IRS of $B(2,1)$ with geometry are discussed.






\section{The group $B^{2,1}(N^{+})$}
\label{Group}

\indent

We turn now to the study of  to $B(2,1)$, the analogue of $B$ in three space$-$time dimensions.
\ Recall that the $2+1$
Minkowski space is the vector space $R^{3}$ of row vectors with 3
real components, with inner product defined as follows. \ Let
$x,y\in R^{3}$ have components $x^{\mu }$ and $y^{\mu }$
respectively, where $\mu =0,1,2$. \ Define the inner product
$x.y$ between $x$ and $y$ by
\begin{equation}
x.y=x^{0}y^{0}-x^{1}y^{1}-x^{2}y^{2}.
\end{equation}
Then the $2+1$ Minkowski space, sometimes written
$R^{2,1}$, is just $R^{3}$ with this inner product.
The ``2,1'' refers to the one plus and two minus signs in the inner product.
Let $SO(2,1)$ be the (connected component of the identity element of the)
group of linear transformations preserving the inner product. \ Matrices
$\Lambda \in SO(2,1)$ are taken as acting by matrix multiplication from the
right, $x\longmapsto x\Lambda ,$ on row vectors $x\in R^{2,1}$. \

\indent
The future null cone $ N^{+} \subset R^{2,1}$ is just the set of nonzero vectors
with zero length and $x^{0}>0$:
\begin{equation}
 N^{+}=\left\{ x\in R^{2,1}| x \cdot x=0, x^{0}>0
\right\} .
\end{equation}
Let $R^{+}$  denote the multiplicative group of all
\it{positive}
\normalfont
real numbers. \ Obviously, if $x\in N^{+}$, then
$tx\in N^{+}$ for any $
t\in R^{+}$. \ Let $F_{1}(N^{+})$ denote the vector space (under
pointwise addition) of all functions $f:N^{+}\rightarrow R$ satisfying
the homogeneity condition
\begin{equation}
\label{hom}
f(tx)=tf(x)
\end{equation}
for all $x\in N^{+}$ and $t\in R^{+}$. \ Define a representation $T$
of $SO(2,1)$ on $F_{1}(N^{+})$ by setting, for each $x\in N^{+}$ and $\Lambda \in SO(2,1)$,
\begin{equation}
(T(\Lambda )f)(x)=f(x\Lambda ).
\end{equation}

\indent
Now let $B^{2,1}(N^{+})$ be the semi$-$direct product
\begin{equation}
\label{tvyrikolasexy}
B^{2,1}(N^{+})=F_{1}(N^{+})
\circledS _{T}
SO(2,1).
\end{equation}
That is to say, $B^{2,1}(N^{+})$ is, as a set, just the product
$F_{1}(N^{+})\times
SO(2,1)$, and the group multiplication law for pairs is
\begin{equation}
(f_{1},\Lambda _{1})(f_{2},\Lambda _{2})=(f_{1}+T(\Lambda _{1})f_{2},\Lambda
_{1}\Lambda _{2}).
\end{equation}



\section{The double cover $B^{2,1}(N^{+})_{c}$}
\label{dc}

\indent

Let ${\rm S}{\rm L}(2,R)$ be the group of all real $2\times 2$ matrices with determinant one.
${\rm S}{\rm L}(2,R)$ is sometimes denoted $G$ below.
Let ${\rm M}_{s}(2,R)$ be the set of all $2\times 2$ symmetric real matrices.
We define a right action of $G $ on ${\rm
M}_{s}(2,R)$ by ${\rm M}_{s}(2,R) \times G \rightarrow {\rm M}_{s}(2,R)$
with \be \label{axcsdfvbgtre}(m,g)\mapsto g^{\top}mg, \ee
where the superscript ${\top}$ means transpose.
Clearly any element $\mu \in {\rm M}_{s}(2,R)$ can be parameterized as follows
$$
\mu=\left[
\begin{array}{cc}
   x^{\rm o}-x^{1}    &   x^{2} \\
    x^{2}     &  x^{o}+x^{1}
\end{array}  \right]
$$
where $x^{\rm o},x^{1},x^{2} \in R$.

\indent
We now consider
the map $b:R^{3} \rightarrow {\rm M}_{s}(2,R)$ defined by \be
\label{wyujhdcvxsertzawl} b(x)=\left[
\begin{array}{cc}
   x^{\rm o}-x^{1}    &  x^{2} \\
    x^{2}     &  x^{o}+x^{1}
\end{array}  \right],
\ee where the $x^{\mu}$ are the components of $x\in R^{3}$. This
map is a linear bijection, so the right action of $G$ on ${\rm
M}_{s}(2,R)$ induces a linear right action of $G$ on $R^{3}$.
Since \be \det(b(x))=x \cdot x \ee and the $G$ action
preserves determinants (indeed $\det g= 1$) in ${\rm
M}_{s}(2,R)$, $G$ acts as transformations from $SO(2,1)$. In fact,
this construction gives an homomorphism \be \gamma : G
\rightarrow SO(2,1)
\ee
which is onto, and has kernel $ Z_{2}=\{ Id,-Id \} $ in
$G$, $Id$ denotes the identity element of $G$.

\indent
Thus $ \gamma $ identifies $ G  $ as the
double cover of $ SO(2,1) $ \be G = SO(2,1)_{c}. \ee
Therefore, the double cover of the group $B^{2,1}(N^{+})$,
given in (\ref{tvyrikolasexy}), has the form
\be B^{2,1}(N^{+})_{c}=
F_{1}(N^{+}) \circledS _{T} G.
\ee
Strictly speaking, $``T"$ should
read $``T\gamma"$, but the notation is simpler as above.

\section{
The group $B^{2,1}(N^{+})_{c}$
}
\label{closeup}

\indent

We
 describe now the structure of the group
$B^{2,1}(N^{+})_{c}.$

\subsection{The projective space of $N^{+}$
}
\label{projnc}


\indent


The null cone $N^{+}$ becomes, under identification (\ref{wyujhdcvxsertzawl}),
$\widetilde{N^{+}}=b(N^{+})$ with

\begin{equation}
\widetilde{N^{+}}
=\left\{ \mu \in
{\rm M}_{s}(2,R),\;\mid x^{0} > 0,   \;
\det \mu=0\right\} .
\end{equation}
Thus $\mu \in \widetilde{N^{+}}
$ if and only
if $\mu $ has rank exactly $1$ and $x^{0} > 0$.
Let $\scri$ be the set of all non$-$zero real two$-$component row vectors $\sigma$; $\scri=R^{2}- 0$.
$\scri$ is
the real section of
Penrose's ``spin space"
appropriate to $N^{+}   \subset R^{2,1}$.
From the rank condition and the requirement $x^{0} > 0$, it follows that
\begin{equation}
\label{sp}
\mu \in \widetilde{N^{+}}
\quad \Leftrightarrow \quad \mu =\sigma^{\top}\sigma
\end{equation}
where  $\sigma \in \scri$.
\noindent
Equation (\ref{sp}) gives a parametrisation of
the null cone $
\widetilde{ N^{+}}
$ by
means  of
the non$-$zero real two$-$component row vectors $\sigma$.
However, this parametrisation, though (by construction) onto, is not one$-$one.

To make this explicit
we define an equivalence relation in $\scri$.
 Two elements  $\sigma,  \sigma^{\prime}$ of  $\scri$ are equivalent if and only if
\be
\label{er}
\sigma^{\prime} = \kappa \sigma, \ \kappa=1,-1.
\ee
Therefore  if
$[\sigma]$ is the equivalence class of $\sigma$,
$[\sigma]$= $\{ \sigma, -  \sigma \}$.

If $\sigma^{\prime}$ and $\sigma$
belong to the same equivalence class,
then, by Eq. (\ref{er}),
$\sigma^{\prime} = \kappa \sigma, \ \kappa=1,-1$.
Therefore,
$$\sigma^{\prime \top} \sigma^{\prime}=  (\kappa \sigma)^{\top} (\kappa   \sigma)=\kappa^{2} \sigma^{\top}  \sigma=\sigma^{\top}  \sigma .$$
Consequently, $\sigma^{\prime}$ and $\sigma$
define the same $\mu.$

Conversely,
if
\noindent
$\sigma^{\prime}$ and $\sigma$
define the same $\mu$, then
\be
\label{rel}
\sigma^{\prime \top} \sigma^{\prime}=\sigma^{ \top} \sigma.
\ee
Postmultiplying (\ref{rel})   by the transpose $\sigma^{ \top}$ of $\sigma$ gives
\be \label{rel1} \sigma^{\prime \top} (\sigma^{\prime}\sigma^{\top})=\sigma^{ \top} (\sigma \sigma^{\top} ). \ee
Since $\sigma \sigma^{\top} \neq 0$, $\sigma^{\prime} \sigma^{\top}$ cannot be
zero.
Hence $\sigma^{\prime}$ is a non zero real multiple of $\sigma$, $\sigma^{\prime}=\lambda \sigma$. Substituting this into
equation
(\ref{rel1}) gives $\lambda^{2}=1$.
Consequently, $\sigma^{\prime}$ and $\sigma$ belong to the same equivalent class.
We therefore arrive at the Proposition which follows.

\begin{prop}
\label{p1}
Every $\mu \in \widetilde{N^{+}}$   corresponds exactly to
one equivalence class $[\sigma]$ and vice versa.
\end{prop}


Let $\pi$ be the projection $\pi: \scri \rightarrow \widetilde{ N^{+}}$ given by
\be
\label{projection}
\pi(\sigma)=\sigma^{\top}\sigma.
\ee
Let $Z_{2}=\{-1, 1\}$.
Define a  left action $Z_{2} \times \scri \rightarrow \scri$ by
\be
\label{map}
(\kappa,\sigma) \mapsto \kappa \sigma.
\ee
Then this $Z_{2}$ action is fixed point free. Indeed
Let $\sigma=(\sigma_{1} \  \sigma_{2}) \in \scri$. If $\lambda \sigma=\sigma$, then,
$\lambda \sigma_{1}  = \sigma_{1}$
and $\lambda \sigma_{2}  = \sigma_{2}$. But $\sigma_{1}$ and $\sigma_{2}$ do not both vanish, so $\lambda=1$ and the action (\ref{map}) is fixed point
free.

\indent
According to  Proposition \ref{p1}
the $Z_{2}$ orbits are precisely the fibres of the projection
\be
\label{proj}
\pi: \scri \rightarrow \widetilde{ N^{+}}.
\ee
We conclude that

\begin{prop}
\label{z2action}
The left free action of $Z_{2}$  on  $\scri$
$$
Z_{2} \times \scri \rightarrow \scri, \qquad \left ( \kappa, \sigma   \right ) \mapsto  \kappa \sigma,
$$
gives rise to a principal $Z_{2}-$bundle $\left (  \scri, \pi,   \widetilde{ N^{+}}  \right )$
with projection
$$
\pi: \scri \rightarrow \widetilde{ N^{+}}, \qquad \pi(\sigma)=\sigma^{\top}\sigma,
$$
and base space diffeomorphic to $\widetilde{ N^{+}}$.
\end{prop}

Let $R^{+}$  be the multiplicative group of positive real numbers.
Define a left action $R^{+} \times \widetilde{ N^{+}} \rightarrow \widetilde{ N^{+}}$ by
\be
(t,\mu) \mapsto t\mu.
\ee
This action is fixed point free and defines a principal bundle $\widetilde{N^{+}} \rightarrow P(\widetilde{N^{+}})$.


\indent
Combining the projections
\be
\scri \rightarrow \widetilde{N^{+}} \rightarrow P(\widetilde{N^{+}})
\ee
evidently corresponds to combining the $Z_{2}$ and $R^{+}$  actions to give  a $R^{+} \times Z_{2}$ action on $\scri$.
Indeed,
we can take the combined action as
\be
\label{pr2}
((t,\lambda),\sigma) \mapsto t \lambda \sigma.
\ee
But, $t\lambda=\kappa \in R^{*}$, where $R^{*}$ is the multiplicative group of non$-$zero real numbers.

Consequently, the
action  (\ref{pr2})     is equivalent to the $R^{*}$ action given by
\be
(\kappa,\sigma) \mapsto \kappa \sigma.
\ee
Hence the base $
P(\widetilde{ N^{+}})$ is the set of orbits for this last action, namely

\be
\label{prnullcone}
P(\widetilde{ N^{+}})  \simeq S^{1}/Z_{2} \equiv P_{1}(R).
\ee
Here $P_{1}(R)$ is the one$-$dimensional real projective space.
Hence we have
\begin{prop}
\label{proj}
The left free action of $R^{*}$  on  $\scri$
$$
R^{*} \times \scri \rightarrow \scri, \qquad \left ( \kappa, \sigma   \right ) \mapsto  \kappa \sigma,
$$
gives rise to a principal $R^{*}-$bundle $\left (  \scri, \pi,   P_{1}(R)  \right )$
with projection
\begin{eqnarray}
\pi   & : & \scri \rightarrow P_{1}(R),  \nonumber \\
\pi(\sigma)& = & \pi(\sigma_{1} \  \sigma_{2})=     \left (     \frac{       \rho}{w(\rho)},   \frac{1}{w(\rho)} \right ),
\nonumber \\
&  &  \rho=\frac{\sigma_{1}}{\sigma_{2}}, \ w(\rho)=(\rho^{2}+1)^{1/2},                     \nonumber \\
&&    when \ \sigma_{2} \neq 0, \ and,     \nonumber    \\
\pi(\sigma)& = & \pi(\sigma_{1} \  0)= (1 \  0), when \ \sigma_{2} = 0,
\end{eqnarray}
and base space diffeomorphic to $P_{1}(R)$.
\end{prop}
\noindent
Noticing that
$\widetilde{N^{+}}=b(N^{+})$ we also have
\be
  P( N^{+}) \simeq  P(\widetilde{ N^{+}})   \simeq P_{1}(R).
\ee

\subsection{The bijections $F(P(N^{+}))      \leftrightarrow F_{1}(N^{+}) \leftrightarrow F_{1}( \widetilde{ N^{+}}) \leftrightarrow F_{2}(\scri)  $}


\indent

Let
\be
\label{hom1}
F_{1}(\widetilde{ N^{+}})=   \{ \tilde{f}: \widetilde{ N^{+}} \rightarrow R \quad | \quad   \tilde{f}(t \mu)=t\tilde{f}(\mu)  \}
\ee
for all $t \in R^{+}$ and $\mu \in \widetilde{ N^{+}}.$

We have the following

\begin{prop}
\label{prop3a}
There are the bijections

$$F(P(N^{+}))      \leftrightarrow F_{1}(N^{+}) \leftrightarrow F_{1}( \widetilde{ N^{+}}).$$
\end{prop}
\begin{prf}
The orbits of the dilatation action $R^{+} \times N^{+} \rightarrow N^{+}; \ (t,x) \mapsto tx$
are
open half$-$lines in $N^{+}$ from the origin of $R^{2,1}$. The projective
null cone
$P( N^{+})$
is the space of these half lines.
Since the homogeneity constraint (\ref{hom}) fixes the
behaviour of functions $f \in F_{1}(  N^{+})$
along
these half lines,
$F_{1}(N^{+})$  may also be realized as the set $F(P(N^{+}))$ of
all arbitrary (unconstrained or ``free'')
real valued functions
defined on $P( N^{+})$:
\be
\label{bi1}
F_{1}(N^{+}) \simeq  F(P(N^{+})).
\ee




Let $x=(x^{1},x^{2},x^{0}) \in N^{+}$ and let the unit circle
$$
S^{1}_{m}= \{ x \in N^{+} \quad | \quad x^{0} = 1    \}
$$
be our model for $P( N^{+})$.
The orbit $ R^{+} x$
intersects $S^{1}_{m}$ at exactly one point
\be
\label{unit}
\tilde{x}=
\frac{x}
{\sqrt{(x^{1})^{2}+(x^{2})^{2}}}.
\ee

With each  $f \in F_{1}(N^{+})$ we associate a function $\alpha \in F(P(N^{+}))$ by restricting $f$
to the circle $S^{1}_{m}$:
\be
\alpha(\tilde{x})=f(\tilde{x}).
\ee
Using the homogeneity  condition  (\ref{hom}) we have
%
\be
\label{cor}
\alpha(\tilde{x})=\frac{f(x)}{\sqrt{(x^{1})^{2}+(x^{2})^{2}}}.
\ee
Formula (\ref{cor}) associates not only $f$ with the corresponding $\alpha$, but also vice versa.
It explicitly gives the bijection
\be
 F_{1}(N^{+}) \leftrightarrow F(P(N^{+})).
\ee

\indent
Evidently, the bijection between $ F_{1}(N^{+})$ and $F_{1}( \widetilde{ N^{+}})$
is given by
\be
f(x)=\widetilde{f}(b(x)).
\ee
This completes the proof.
\end{prf}

\noindent

Now suppose that $\psi: \scri \rightarrow R$ is any function which satisfies
\be
\label{even}
\psi(-\sigma)=\psi(\sigma)
\ee
for all $\sigma \in \scri$. This just means that $\psi$ is constant on each $Z_{2}$ orbit of Proposition \ref{z2action}.
That is, $\psi$ only depends on the points $\mu=\sigma^{\top}\sigma \in \widetilde{ N^{+}}$, and so can be written as a function
$\tilde{f}:\widetilde{ N^{+}} \rightarrow R:$
\be
\label{psi}
\psi(\sigma)=\tilde{f}(\sigma^{\top}\sigma).
\ee
Conversely, any function $\tilde{f}: \widetilde{ N^{+}} \rightarrow R$ defines a function $\psi$, given by (\ref{psi}), which satisfies (\ref{even}).


We can now use (\ref{psi}) to identify $F_{1}( \widetilde{ N^{+}})$
with a subset of the space
of all functions $\psi:\scri \rightarrow R$. Let $F_{2}(\scri)$ be the set of all such functions which are homogeneous of
degree 2, i.e., of all functions satisfying
\be
\label{hom2}
\psi(q \sigma)= q^{2} \psi(\sigma)
\ee
for all $q \in R^{*}$ and all $\sigma \in \scri$. Then we have the following
\begin{prop}
\label{prop4}
There is a bijection
$$F_{1}( \widetilde{ N^{+}})      \leftrightarrow F_{2}(\scri).$$
\end{prop}
\begin{prf}
Given any $\psi \in F_{2}(\scri)$, we have, using (\ref{hom2}), $\psi(-\sigma)=\psi(\sigma)$, so $\psi$
satisfies (\ref{even}). So, as in (\ref{psi}), we can write $\psi(\sigma)=\tilde{f}(\sigma^{\top}\sigma)$
for some function $\tilde{f}:\widetilde{ N^{+}} \rightarrow R.$ Writing $\mu=\sigma^{\top}\sigma$, we then have
$$
\tilde{f}(t\mu)=\tilde{f}(t(\sigma^{\top}\sigma))=\tilde{f}((\sqrt{t}\sigma)^{\top}(\sqrt{t}\sigma))=\psi(\sqrt{t}\sigma)=
t\psi(\sigma)=t\tilde{f}(\sigma^{\top}\sigma)=t \tilde{f}(\mu),
$$
for all $t \in R^{+}$, so $\tilde{f} \in F_{1}( \widetilde{ N^{+}}) $.

\noindent
Conversely, given any $\tilde{f} \in F_{1}( \widetilde{ N^{+}})$, define $\psi(\sigma)=\tilde{f}(\sigma^{\top}\sigma)$.
 Then, for any $q \in R^{*}$, we have
\be
\psi(q \sigma)= \tilde{f} ((q \sigma)^{\top}(q \sigma))=\tilde{f}(q^{2}(\sigma^{\top}\sigma))=
q^{2}\tilde{f}( \sigma^{\top}\sigma))= q^{2} \psi(\sigma).
\ee
Thus $\psi$ satisfies (\ref{hom2}), so $\psi \in F_{2}(\scri).$ This completes the proof.

\end{prf}

\noindent
Propositions (\ref{prop3a}) and (\ref{prop4}) lead to
\begin{prop}
There are the bijections
$$F(P(N^{+}))      \leftrightarrow F_{1}(N^{+}) \leftrightarrow F_{1}( \widetilde{ N^{+}}) \leftrightarrow  F_{2}(\scri).$$
\end{prop}

\subsection{The spinor version of
$B^{2,1}(N^{+})_{c}$}


\indent

The relationship between $g \in G$ and the image $\Lambda \in SO(2,1)$
is given by
\be
\label{hom3}
g^{\top}b(x)g=b(x\Lambda).
\ee
The action (\ref{axcsdfvbgtre}) restricts to an action of $G$ on $\widetilde{ N^{+}}$.
For $x \in N^{+}$,
$b(x)=\sigma^{\top}\sigma$, $\sigma \in \scri$.
Then equation (\ref{hom3}) reads
\be
g^{\top} \sigma^{\top} \sigma g=b(x\Lambda).
\ee
Thus
\be
b(x\Lambda)=(\sigma g)^{\top}(\sigma g).
\ee

Since the identification between $F_{1}(N^{+})$ and $F_{1}( \widetilde{ N^{+}})$ is given by
\be
f(x)=\widetilde{f}(b(x)),
\ee
we may now compute the representation $T(g)$
in terms of $F_{1}( \widetilde{ N^{+}})$ as follows. We have
\be
(T(\gamma(g))f)(x)=
(T(\Lambda)f)(x)=f(x\Lambda)=\tilde{f}(b(x \Lambda))=\tilde{f}((\sigma g)^{\top}(\sigma g))=\psi(\sigma g).
\ee

Now define a new representation $T$ of $G$ in $F_{2}(\scri)$ by the following formula
\be
\label{action1}
(T(g)\psi)(\sigma)=\psi(\sigma g).
\ee
This representation reduces to the identity on the subgroup $\{ Id,-Id \}$ of $G$
(that is on the kernel of the homomorphism from $G$ onto $SO(2,1)$).
We have just seen
that it corresponds to the representation $T(g)$ of $G$ on $F_{1}(  N^{+})$
defining $B^{2,1}(N^{+})_{c}$.
Hence
\begin{thrm}
\label{TH1}
The  spinor version of $B^{2,1}(N^{+})_{c}$
is defined by
\be
B_{s}^{2,1}(N^{+})_{c}=F_{2}(\scri) \circledS _{T} G,
\ee
where
the representation $T$
is given by
$$
(T(g)\psi)(\sigma)=\psi(\sigma g).
$$
\end{thrm}

\subsection{Unconstrained supertranslations}

\indent


In considering representation theory, it is convenient to work in terms of unconstrained functions. To this end we find a new realization of $B_{s}^{2,1}(N^{+})_{c}
$
with $F_{2}(\scri)$ realized as $F(P(N^{+}))$.

First
we introduce convenient local coordinates  into $\scri$.
Given $\sigma=(\sigma_{1},\sigma_{2}) \in  \scri$, assume that  $\sigma_{2}  \neq 0.$
Let $r(\sigma)$ be the length of $\sigma$ and let $\mathcal S (\sigma_{2})$ be the sign of
$\sigma_{2}$:
\be
r(\sigma)=(\sigma_{1}^{2}+\sigma_{2}^{2})^{1/2}, \quad \mathcal S (\sigma_{2})=\frac{\sigma_{2}}{|\sigma_{2}|}.
\ee
Further define
\be
\rho=\frac{\sigma_{1}}{\sigma_{2}}, \quad w=w(\rho)=\frac{r}{|\sigma_{2}|}=(\rho^{2}+1)^{1/2}.
\ee

Thus $\sigma$ can be written as
\be
\sigma=(\sigma_{1},\sigma_{2})=r \left (   \frac{\sigma_{1}}{r},\frac{\sigma_{2}}{r}  \right )=
r \mathcal S (\sigma_{2})    \left (     \frac{       \rho}{w(\rho)}, \frac{1}{w(\rho)} \right ),
\ee
and so, writing the unit vector in the final brackets as $e(\rho)$, we have
\be
\sigma=r(\sigma) \mathcal S (\sigma_{2})  e(\rho).
\ee
The vector $e(\rho)$ has the second component positive. Evidently $\rho$ is a local projective
coordinate for $P(N^{+})$.

Using
the homogeneity condition (\ref{hom2}) we have
\be
\psi(\sigma)=\psi(r(\sigma) \mathcal S (\sigma_{2})  e(\rho))=r^{2}(\sigma)\mathcal S^{2}(\sigma_{2})\psi(e(\rho)).
\ee
Writing $\psi(e(\rho))=\alpha(\rho)$ this expresses (locally) every $\psi \in F_{2}(\scri)$  in terms of a function
$\alpha \in  F(P(N^{+})$ by
\be
\psi(\sigma)= r^{2}(\sigma)\alpha(\rho).
\ee

To find the expression for the $T(g)$ operators in terms of the $\alpha$s note first that if $g \in G$ is
$$
\left[
\begin{array}{cc}
   a    &  b \\
    c     &  d
\end{array}  \right]
$$
then the components $\sigma_{1},\sigma_{2}$ of $\sigma$ transform linearly, so that the ratio $\rho=\frac{\sigma_{1}}{\sigma_{2}}$
transforms fraction linearly. Writing $\rho g$ for the transformed ratio,
\be
\label{actpr}
\rho g= \frac{(\sigma g)_{1}}{(\sigma g)_{2}}=\frac{\sigma_{1}a+\sigma_{2}c}{\sigma_{1}b+\sigma_{2}d}=\frac{\rho a + c }{\rho b + d}.
\ee
Hence we have
\be
\psi(\sigma g)=r^{2}(\sigma g) \alpha(\rho g).
\ee

The action $T$
of $G$ on the $\psi$s in Theorem \ref{TH1} induces an action, also denoted $T$, of $G$
on the $\alpha$s defined by
\be
(T(g)\psi)(\sigma)=r^{2}(\sigma)(T(g)\alpha)(\rho).
\ee
The last two equations give
\be
\label{actb}
(T(g)\alpha)(\rho)=\kappa_{g}(\rho)\alpha(\rho g),
\ee
where the factor $\kappa_{g}(\rho)$ on the right is defined by
\be
\label{kfac}
\kappa_{g}(\rho)=\frac{r^{2}(\sigma g)}{r^{2}(\sigma)}=\frac{(a \sigma_{1}+c\sigma_{2} )^{2}+(b \sigma_{1}+d\sigma_{2})^{2}}{\sigma_{1}^{2}+\sigma_{2}^{2}}=\frac{(\rho a + c)^{2}+(\rho b + d)^{2}}{\rho^{2}+1}.
\ee
Note that the last two formulae are expressed entirely in terms of the local coordinate $\rho$ for $P_{1}(R)$. Strictly
speaking one more local chart is needed to cover all of $P_{1}(R)$ related to the above via $\frac{1}{\rho}$, but the
single one used above will be sufficient here.

Summarising, we have
\begin{thrm}
\label{th2}
The group $B_{s}^{2,1}(N^{+})_{c}
$ can be realized as
\be
\label{defb}
B_{s}^{2,1}(N^{+})_{c}=F(P(N^{+})) \circledS _{T} G
\ee
with semidirect product specified by
\be
(T(g)\alpha)(\rho)=\kappa_{g}(\rho)\alpha(\rho g).
\ee
\end{thrm}


\subsection{Almost unconstrained supertranslations}

\indent

We now find a new model for $F(P(N^{+}))$ on the double cover
of $P_{1}(R)$. With each vector $\sigma=(\sigma_{1},\sigma_{2})  \in  \scri$
we associate the unit length vector $  e_{\sigma} =
\sigma/ |\sigma| $
having the same direction as $\sigma.$ Thus we have
\be
\label{cord}
\sigma= r(\sigma) e_{\sigma}, \quad   r(\sigma) = |\sigma|, \quad   e_{\sigma} =
\sigma/ |\sigma|.
\ee

Let $S^{1} \subset \scri$ be the set of vectors of unit length in $\scri$:
\be
S^{1}_{m}= \{ \sigma \in \scri \quad |  \quad     |\sigma| = 1    \}.
\ee
Let $F_{e}(S^{1})$ be the set of all real valued $even$ functions $\beta:S^{1} \rightarrow R.$
That is these functions satisfy the even$-$ness condition
\be
\beta(-e_{\sigma})=\beta(e_{\sigma}).
\ee

With each  $\psi \in F_{2}(\scri)$ we associate a function $\beta \in F_{e}(S^{1})$ by restricting $\psi $
to the circle $S^{1}$:
\be
\label{res}
   \beta(e_{\sigma})=\psi(e_{\sigma}).
\ee
Using the homogeneity  condition  (\ref{hom2}), we have
\be
\psi(\sigma)=\psi( r(\sigma)  e_{\sigma}   ) = r^{2}(\sigma) \psi(   e_{\sigma} ) = r^{2}(\sigma)    \beta(   e_{\sigma} ).
\ee
Hence we can write
\be
\label{cor}
\beta(   e_{\sigma} )= \frac{    \psi(\sigma)                }    {    r^{2}(\sigma)      }.
\ee
Formula (\ref{cor}) associates not only any $ \psi$ with the
corresponding $\beta$, but also vice versa. So we have a
bijection
\be
F_{2}(\scri) \leftrightarrow F_{e}(S^{1}),
\ee
giving us a new realisation of the supertranslation space $F_{2}(\scri).$
In this new realisation, the supertranslations are ``almost unconstrained'',
they are subject to the condition that they must be even.

Hence we can now calculate the representation $T(g)$ in this new
realisation $F_{e}(S^{1})$. The transformation $\sigma \mapsto \sigma^{\prime} = \sigma g$ becomes,  in the coordinates defined by (\ref{cord}),
$r(\sigma) \mapsto  r(\sigma^{\prime}) $ and
$e_{\sigma}  \mapsto  e_{\sigma^{\prime}}   $ where
\begin{eqnarray}
\label{coef1}
r(\sigma^{\prime})&=&|\sigma^{\prime}|=|\sigma g|=|r(\sigma) e_{\sigma}g|=r(\sigma)|e_{\sigma}g|, \quad \rm{and}, \\
\label{coef2}
e_{\sigma^{\prime}}&=&e_{\sigma g}=(\sigma g)/|\sigma g|=(r(\sigma) e_{\sigma}  g  )/ |r(\sigma) e_{\sigma}  g   | = (e_{\sigma}  g)/|e_{\sigma}  g|.
\end{eqnarray}

Let us denote $T(g)\psi$ by $\psi^{\prime}$, and, correspondingly,  $T(g) \beta$ by $\beta^{\prime}$. Then,
using  (\ref{action1}),
(\ref{res}),   (\ref{cor}), and  (\ref{coef1}),
(\ref{coef2}), we get
\begin{eqnarray}
(T(g)\beta)(e_{\sigma}) &  =  & \beta^{\prime}(e_{\sigma})=
\frac{\psi^{\prime}(\sigma)}{ r^{2}(\sigma)}=
\frac{\psi(\sigma g)}{ r^{2}(\sigma)}=
\frac{\psi(\sigma^{\prime})}{ r^{2}(\sigma)}=
\frac{\psi(  r(\sigma^{\prime})    e_{\sigma^{\prime}}   )}{ r^{2}(\sigma)}=\frac{  r^{2}(\sigma^{\prime})      }{ r^{2}(\sigma)}
\psi(     e_{\sigma^{\prime}} )    \nonumber \\
&=&
\frac{  r^{2}(\sigma^{\prime})      }{ r^{2}(\sigma)}
\beta(     e_{\sigma^{\prime}} )=
\kappa_{g}(e_{\sigma})
\beta(     e_{\sigma^{\prime}} )=
\kappa_{g}(e_{\sigma})
\beta(     e_{\sigma g} ),
\end{eqnarray}
where the factor $\kappa_{g}(e_{\sigma})$ on the right
is defined by
\be
\label{kfac1}
\kappa_{g}(e_{\sigma})=\frac{  r^{2}(\sigma^{\prime})      }{ r^{2}(\sigma)} =  \frac{|\sigma g|^{2}}{|\sigma |^{2}}=
|e_{\sigma}g|^{2}.
\ee
From (\ref{kfac}) and (\ref{kfac1}) we have
\be
\kappa_{g}(e_{\sigma})=\kappa_{g}(\rho).
\ee

We conclude that
\begin{thrm}
\label{th3}
The group $B_{s}^{2,1}(N^{+})_{c}
$ can be realized as
\be
\label{defbe}
B_{s}^{2,1}(N^{+})_{c}=
F_{e}(S^{1})
\circledS _{T} G
\ee
with semidirect product given by
\be
\label{actbe}
(T(g)\beta)(e_{\sigma})=\kappa_{g}(e_{\sigma})\beta(e_{\sigma g}).
\ee
\end{thrm}
We can use either of the realizations
given in Theorems
\ref{th2} and \ref{th3} to develop the
representation theory of $B_{s}^{2,1}(N^{+})_{c}$.
Here we
use
the realization
given in Theorem \ref{th2}.

\section{The group $B(2,1)$}
\label{anB}

\indent

So far, the supertranslation space $F(P(N^{+}))$  has been defined as a space of truly arbitrary functions. This has been
merely for clarity; for physical applications it is necessary to give to this space additional structure.
To this end we now restrict $F(P(N^{+}))$ to be the Hilbert space of square integrable functions on
$P(N^{+})
$ with respect to the standard (Lebesgue) measure.

It will be convenient then to use as coordinates for $P(N^{+})$
the two components of a unit length vector $\textbf{m}$  in $R^{2}$
 \be
P(N^{+})=
\{ \textbf{m}   \in  R^{\rm{2}}| \ | \textbf{m} \normalfont | =  \rm{1}       \}.
\ee
The
$\textbf{m}$ \normalfont coordinates are globally defined.
The usual angular coordinate $\theta$ for $P(N^{+})$ is defined by
\be
\textbf{m}=(\rm{sin\theta},\rm{cos\theta})=(\textbf{m}_{\rm 1},\textbf{m}_{\rm 2}),
\ee
$\theta \in [0,2 \pi)$.
The  angular coordinate $\theta$ is related  to the projective coordinate $\rho$ for $P(N^{+})$
by stereographic projection
\be
\rho=\cot \left(  \frac{\theta}{2}  \right).
\ee

 The required measure
$\lambda$ on $P(N^{+})$ is the usual
(normalized) length of an arc segment for $S^{1}$. The explicit form is the 1$-$form
\be
d \lambda(\textbf{m})=\frac{\rm{d}\textbf{m}_{\rm 1}}{\rm{2} \pi \textbf{m}_{\rm 2} },
\ee
where $\textbf{m}_{\rm 1},\textbf{m}_{\rm 2}$ are the components of a unit vector $\textbf{m} \in$$ \ R^{2}$. This formula is only valid for
$\textbf{m}_{\rm 2} \neq \rm{0}$, but a permutation of the indices in the components gives another valid expression for $d \lambda(\textbf{m})$.
In terms of $\theta$
the normalized Lebesgue measure is given by
\be
\label{mes1}
d \lambda(\theta
)=\frac{\rm{d} \theta}{\rm{2} \pi}
\ee
which
evidently is $SO(2)-$invariant.
In terms of
$\rho$
the corresponding expression is
\be
\label{mes2}
d \lambda(\rho)=-\frac{d \rho}{\pi(1+\rho^{2})}.
\ee

Letting $\textbf{m}$$g$ be the transform of $\textbf{m}$
by $g \in G$ (so that $\rho \mapsto \rho g$
corresponds to $\textbf{m} \mapsto \textbf{m}$$g$), a simple calculation gives
\be
\label{ratio}
\frac{  \rm{d} \lambda (\textbf{m}) } {   \rm{d} \lambda ( \textbf{m} \textit{g})}=\kappa_{g}(\rho)=\kappa_{g}(\textbf{m}).
\ee
Here $\kappa_{g}(\textbf{m})$ denotes $\kappa_{g}(\rho)$, where $\textbf{m}$ corresponds to $\rho$. While the expression given
for $\rm{d} \lambda ( \textbf{m})$ is local, $\kappa_{g}(\textbf{m})$ is defined globally.



\noindent

Let $L^{2}(P_{1}(R),\lambda,R)$
be  the space of all functions $\alpha:P_{1}(R) \rightarrow R$
which are Lebesgue measurable and satisfy
\be
\int_{P_{1}(R)} |\alpha(\rho)|^{2} \rm{d}\lambda(\rho)<+\infty.
\ee
Introduce an inner product into $L^{2}(P_{1}(R),
\lambda,R)$ by setting, for each $\alpha_{1},\alpha_{2} \in L^{2}(P_{1}(R),\lambda,R)$,
\be
\label{inpr}
<\alpha_{1},\alpha_{2}>=\int_{P_{1}(R)}   \alpha_{1}(\rho)      \alpha_{2}(\rho)           \rm{d}\lambda(\rho).
\ee
Then, as is well known, $L^{2}(P_{1}(R),\lambda,R)$ becomes a real Hilbert space with respect to this inner product in the usual way.
We denote the resulting  Hilbert space by $\mathcal H (P_{1}(R)).$
In $\mathcal H (P_{1}(R))$
functions are identified whenever they differ, at most, on a set of measure zero.
We use the homeomorphism $P(N^{+})\simeq P_{1}(R)$ to identify the spaces
$L^{2}(P(N^{+}),  \lambda, R)$   and $L^{2}(P_{1}(R), \lambda, R )$,
and consequently,
the ensuing Hilbert spaces   $\mathcal H (P(N^{+}))$  and   $\mathcal H (P_{1}(R))$ respectively.


We now replace in the definition (\ref{defb}) the supertranslation space $F(P(N^{+}))$ by
$\mathcal H (P_{1}(R))$.
Note that the representation $T$, given by (\ref{actb}), restricts to the subspace $\mathcal H (P_{1}(R))$
and it is given by the
same formula as in (\ref{actb}).
So, our final definition of our group, which we will, now and henceforth, denote by $B(2,1)$ (to distinguish it from previous definitions),
is the group
\begin{de}
\be
\label{de}
B(2,1)=\mathcal H (P_{1}(R))
\circledS _{T} G.
\ee
with semidirect product specified by
$$
(T(g)\alpha)(\rho)=\kappa_{g}(\rho)\alpha(\rho g).
$$
\end{de}

We can also use the bijections between all the different models for supertranslation space to restrict all of them
to functions corresponding to $L^{2}(P_{1}(R),\lambda,R)$  and to define
corresponding Hilbert space structures for all of them. Denoting the resulting spaces by $\mathcal H$, and using subscripts, as before,
to indicate constraints, we now have
\be
\mathcal H (P(N^{+}))      \leftrightarrow \mathcal H_{1}(N^{+}) \leftrightarrow \mathcal H_{1}( \widetilde{ N^{+}}) \leftrightarrow \mathcal H_{2}(\scri).
\ee



We note
that there are two distinct transitive actions of $G$ on $S^{1}$.
\begin{enumerate}

\item{ The action (\ref{actpr}) on $P_{1}(R)$ induced by
the usual linear action of $G$ on $R^{2}$.
The stabilizer of $\infty$ is the
subgroup $\mathcal U$ of $G$ of upper triangular matrices. Therefore,
$P_{1}(R)$ is in bijective correspondence with and homeomorphic to the quotient space $G/\mathcal U.$ In this
case the action (\ref{actpr}) of $G$ on $P_{1}(R)$ is not faithful; the kernel of the action is the
group $\left \{  \begin{pmatrix}
1 & 0 \\
0 & 1
\end{pmatrix}
, \
\begin{pmatrix}
-1 & \; \; \; 0 \\
\; \; \; 0 & -1
\end{pmatrix}  \right  \}  \simeq   Z_{2}.
$ This action was used to define the representation $T$ in (\ref{actb}) which specifies the semi$-$direct product
$B(2,1)$ in (\ref{de}).
}

\item{The action on the circle $ \left (  R^{2} - \{  0   \}    \right )  /  R_{>0}$ inferred from (\ref{coef2}) is
\be \label{action1}  e_{\sigma}g=e_{\sigma^{\prime}}=(e_{\sigma}  g)/|e_{\sigma}  g|.\ee
If $e_{\sigma}=(x \ y)$,  $x$ and $y$ are the cartesian coordinates of $e_{\sigma},$ action (\ref{action1}) reads
\be(x \ y)
\begin{pmatrix}
a &  b \\
c & d
\end{pmatrix} =\frac{1}{ \sqrt{ \left ( a x + c y   \right )^{2}   +    \left ( b x + d y   \right )^{2} }} \left ( a x + c y \  b x + d y    \right ).
\ee
The stabilizer of $(1 \ 0 ) $ for this action is $\mathcal U^{+}$, the subgroup of index 2 of $\mathcal U$  with positive entries in the main diagonal.
Therefore the circle $ \left (  R^{2} - \{  0   \}    \right )  /  R_{>0}$ is in bijective correspondence
with and homeomorphic to the quotient space $G/\mathcal U^{+}.$  In this case the action of $G$ on $G/\mathcal U^{+}$ is faithful.
This action was used to define the representation $T$ in (\ref{actbe}) which specifies the semi$-$direct product
$B_{s}^{2,1}(N^{+})_{c}$ in (\ref{defbe}).}
\end{enumerate}

\noindent
There is  a continuous $G$$-$equivariant surjection $ G/\mathcal U^{+} \longrightarrow G/\mathcal U$   which is a connected 2$-$fold covering.



\section{$B(2,1)$ as the automorphism group of the strong conformal geometry of $\scri^{+}$}
\label{auto}


According to Klein's Erlangen programme, one may (indirectly) specify a geometry by giving a group action. Conversely,
given a group action, one may ask for the corresponding geometry. Penrose
\cite{Pen1}, \cite{Pen2}, \cite{Pen3} interpreted the original BMS group $B$ as a transformation group $B \times \scri^{+} \longrightarrow \scri^{+}$ of the ``future null infinity''
of conformally
compactified weakly asymptotically simple space$-$times.
Furthermore, he gave a geometric structure to $\scri^{+}$, the ``strong conformal geometry'', such that this
classical action  $B \times \scri^{+} \longrightarrow \scri^{+}$ is the group of automorphisms of the geometry. In this section we rederive $B(2,1)$ as the
automorphism group of the ``strong conformal geometry'' of the ``future null infinity'' $\scri^{+}$ of 3$-$dim Minkowski
space$-$time.

In dealing with
$B(2,1)$ it has been convenient to specify the action $T$ of $G$ on
$\scri$
from the right. In order to
agree with notations used elsewhere (e.g. \cite{Pen2}, \cite{Pen3}),
it now becomes convenient to use the action of
$G$ on  $\scri$
from the left. So, for this and the next section only, we redefine the groups and quantities needed.
Let $g \in  G$ and write
\be
\label{eqleft1}
g=\left[
\begin{array}{cc}
   a    &  b \\
    c     &  d
\end{array}  \right], \qquad \sigma=\left[
\begin{array}{c}
  \sigma_{1}     \\
    \sigma_{2}
\end{array}  \right], \qquad g \sigma = \left[
\begin{array}{cc}
   a    &  b \\
    c     &  d
\end{array}  \right] \left[
\begin{array}{c}
  \sigma_{1}     \\
    \sigma_{2}
\end{array}  \right],
\ee
\be
\label{eqleft2}
\rho=\frac{\sigma_{1}}{\sigma_{2}}, \qquad g\rho=\frac{a \rho + b }{c \rho + d}.
\ee

\subsection{``Future null infinity'' $\scri^{+}$  of 3$-$dim Minkowski
space$-$time}

The following definition will be useful

\begin{de}
The 3$-$dim Einstein
universe is a solution to Einstein's equations
with zero cosmological constant in 3 space$-$time dimensions with metric    $\mathcal M$         given by
\begin{eqnarray}
ds^{2}  &= & - dT^{2}+   d\Psi^{2}+sin^{2}\Psi  d\phi^{2},           \\
where,&& -\infty < T < \infty, \ 0 \leq \Psi < \pi, \ 0 \leq \phi  < 2\pi.
\end{eqnarray}
\end{de}

\indent
Einstein's  3$-$dim
universe contains dust with constant proper energy density $\rho$, $\rho=\frac{1}{\kappa},$
$\kappa=\frac{8 \pi G}{c^{4}}.$ It has four killing vector fields
\begin{eqnarray}
\xi_{1}&=&\partial_{T}, \  \xi_{2}=\partial_{\phi},        \     \xi_{3}= - \left ( cos\phi  \partial_{\Psi} - cot \Psi sin \phi \partial_{\phi} \right ),   \\
\xi_{4}&=& sin\phi \partial_{\Psi} + cot \Psi cos \phi \partial_{\phi}.
\end{eqnarray}
$\xi_{2}, \ \xi_{3}, \ \xi_{4}$ are the three killing vector fields of the 2$-$sphere.
$\xi_{1}$ is a global timelike Killing vector field, orthogonal to the level hypersurfaces of $T$ which are 2$-$spheres.
Therefore, 3$-$dim Einstein  universe is static with topology $R \times S^{2}$.
3$-$dim Einstein  universe will be denoted by   $Ein_{3}$,  $Ein_{3}= \left (  R \times S^{2}, \mathcal M         \right ).         $

We obtain
\begin{eqnarray}
\label{con1}
\mathcal M & = & \Omega^{2} \eta, \ \rm where,      \\
\label{con2}
\Omega & = &    \frac{2}     {    \sqrt{(1+u^{2})(1+v^{2})}    },       \\
\label{con3}
\eta   & = &    \frac{(1+u^{2})(1+v^{2})} {4}     \left (    - dT^{2}+   d\Psi^{2}      \right ) +  \frac{  \left (    u   - v     \right )^{2}    }{4}     d\phi^{2}  \nonumber \\
\label{con4}
          & = & -  du   dv +  \frac{  \left (    u   - v     \right )^{2}    }{4}     d\phi^{2}     \nonumber \\
\label{con5}
            & = &   - dt^{2}   +  dr^{2}   + r^{2}    d\phi^{2},      \\
\label{con6}
               &  &   T=arctanu+ arctanv, \ \Psi= arctanv- arctanu, \\
\label{con7}
               &&   u=t-r, \ v= t+r,    \\
\label{con8}
               &&  r \geq 0, \   -\infty < t, \ u, \ v < \infty.
\end{eqnarray}
The metric $\eta $ is the metric of the 3$-$dim Minkowski space$-$time   $Min_{3},$ $Min_{3}=\left (  R^{3}, \eta \right )$.

By using equations (\ref{con1})-(\ref{con8})  we are easily led
to the following
\begin{prop}
\label{con}
There exists a conformal isometry of $Min_{3}$
into the open region
$\mathcal Min_{3}$
of $Ein_{3}$
given by the coordinate restrictions
\be
-\pi < T < \pi, \ 0 \leq \Psi < \pi, \ 0 \leq \phi  < 2\pi.
\ee
\end{prop}

\begin{de}
The conformal infinity of $Min_{3}$
is the boundary
$\mathcal  B_{\mathcal Min_{3}}$
of $\mathcal Min_{3}$ in $Ein_{3}.$
\end{de}
Let $\overline{\mathcal Min_{3}}$ be the closure of
$\mathcal Min_{3}$, i.e.,
$\overline{\mathcal Min_{3}}
=\mathcal Min_{3}  \bigcup  \mathcal  B_{\mathcal Min_{3}}. $

We can easily prove the following
\begin{lm}
The 2$-$dim
surface
$$\scri^{+}= \{  (T, \Psi, \varphi): T=\pi-\Psi, \ 0< \Psi < \pi, \     0 \leq \phi  < 2\pi         \} $$
in $\mathcal  B_{\mathcal Min_{3}}$
is  smooth null hypersurface for $\mathcal M$ (hence the terminology `` null infinity''). Its null generators are the integral curves of the
vector field $ \partial_{T}  -   \partial_{\Psi}. $
\end{lm}

Conformal ``compactification'' is not always possible. The property for a space$-$time to admit
a smooth conformal compactification can be characterized in terms of decay of the Weyl curvature
at infinity. When such a compactification exists
the boundary, which is
 $\mathcal  B_{\mathcal Min_{3}}$ in the case under consideration,
 will have a structure: different
parts corresponding to different ways of going to infinity in the physical space$-$time.
$\scri^{+}$ is the subset of the boundary, i.e., of $\mathcal  B_{\mathcal Min_{3}}$
in our case,
with elements
the end points of all future directed null geodesics of $Min_{3}$
(hence the terminology `` future infinity'').


Conformal ``compactification''  when possible is not unique:
There is a freedom \cite{Pen3}    in the choice of $\Omega$  in (\ref{con1}).
Indeed, $\Omega$ can be replaced by any
\begin{equation}
\Omega^{\prime}= \Theta \Omega,
\end{equation}
$\Theta$   being \it  any \normalfont
smooth positive real valued function on
$\overline{\mathcal Min_{3}}$.
As a result of this freedom, a
conformal
class $\mathcal C_{
Min_{3}}$   of  metrics
is defined on $
Min_{3}.$
Thus $\mathcal C_{
Min_{3}}$
consists of all metrics $\Omega^{2} \mathcal M,$
where $\Omega$ is  a smooth positive real valued
function on $\mathcal Min_{3}$, which vanishes on
$\scri^{+}$, and has non$-$zero divergence on
$\scri^{+}.$

By restricting $\mathcal M$ to $\scri^{+}$ we obtain that the induced metric on $\scri^{+}$
is given by
\begin{eqnarray}
\label{inmetr}
dl^{2}  &= & - 0 d\Psi^{2}+sin^{2}\Psi  d\phi^{2},           \\
where,&&
0 \leq \Psi < \pi, \ 0 \leq \phi  < 2\pi.
\end{eqnarray}


We can use the freedom in the choice of $\Omega$ to scale the metric (\ref{inmetr}) so that the
divergence of the null  generators of $\scri^{+}$ vanishes.
In fact one can easily prove the following
\begin{lm}
\label{lemma2}
With a proper
choice of $\Omega$ the induced metric on $\scri^{+}$ can be cast into the form $\mathcal M_{\scri^{+}}$ given by
\begin{eqnarray}
\label{ejisvsh}
dl^{2} &  =   &  0 du^{2}+ d\phi^{2}=0 du^{2} +  \frac{4}{(1+\rho^{2})^{2}}  d \rho^{2}  \\
where,&& \Psi=\frac{\pi}{2}- arctanu, \ -\infty < u < \infty, \\
\rho & = & \frac{cot\phi}{2}, \  -\infty < \rho < \infty.
\end{eqnarray}
\end{lm}
Therefore,  in the $(u,\phi)$ coordinates, the null generators of $\scri^{+}$, affinely parameterized by $u$, are the integrals curves of the
vector field $\frac{\partial}{\partial u}$. The part of the metric (\ref{ejisvsh}) that is transverse to $\frac{\partial}{\partial u}$
is the metric of a circle  with unit radius.
Therefore, $\scri^{+}$ has topology $
R \times S^{1},$   ``$R$'' being the null generators of  $\scri^{+}$,  ``$S^{1}$''
being the circle of unit radius.


$\mathcal C_{
Min_{3}}$ induces, via (\ref{con1}),
a
conformal
class $\mathcal C_{
\scri^{+}}$   of  metrics
on $
\scri^{+}.$
$\mathcal C_{
\scri^{+}}$
consists of all metrics $\frac{1}{\Omega^{2}} \mathcal M_{\scri^{+}},$
where $\Omega$, as in $\mathcal C_{
Min_{3}}$,    is  a smooth positive real valued
function on $\mathcal Min_{3}$, which vanishes on
$\scri^{+}$, and has non$-$zero divergence on
$\scri^{+}.$


\subsection{The Newman$-$Unti group}

The \it local \normalfont orientation$-$preserving conformal maps of the plane onto itself are given by
$$
\zeta \longrightarrow \overline{\zeta}
=f(\zeta)
$$
where $f$ is holomorphic and $\zeta=\sigma_{1}+i\sigma_{2},$ $\sigma_{1}, \ \sigma_{2}$ being Cartesian coordinates on the plane.

However, for a \it global \normalfont map, we would require that both $f$ and its inverse $f^{-1}$ be non$-$singular over the whole plane.
This restricts $f$ to be a linear function
\be
\label{tr}
f(\zeta)=w \zeta +q.
\ee
Therefore, the group (\ref{tr})   of \it global \normalfont orientation$-$preserving conformal maps of the plane onto itself is a four$-$real$-$parameter
group.
Writing $ w=|w|e^{i \vartheta} $ in polar coordinates, we see that
the affine transformations (\ref{tr})
can be viewed as the composition of a rotation, followed by a scaling, followed by a translation.

The linear action (\ref{tr}) induces
a  fraction linear action on $P_{1}(R)$
\be
\label{ingroup}
\rho \longrightarrow \overline{\rho}=g(\rho)=\frac{a \rho+b}{c \rho+d},
\ee
$\rho=\frac{\sigma_{1}}{\sigma_{2}}, \ a,b,c,d \in R.$
The three real ratios $a:b:c:d$ define $g$. We can normalize $a,b,c,d$  by
\be
a d - b  c =1.
\ee
Thus every map $g$ corresponds to an element of $PSL(2,R)$ and vice versa.

One can easily prove the following

\begin{prop}
The group of non$-$reflective motions of $\scri^{+}$ preserving its intrinsic
degenerate metric
\begin{eqnarray}
dl^{2} &  =   &
0 du^{2} +  \frac{4}{(1+\rho^{2})^{2}}  d \rho^{2}  \nonumber \\
where,&&
-\infty < u < \infty,  \  -\infty < \rho < \infty, \nonumber
\end{eqnarray}
up to conformal rescalings is given by
\begin{eqnarray}
\label{der}
\rho \longrightarrow \overline{\rho}&=&\frac{a \rho+b}{c \rho+d},  \\
\label{der1}
u \longrightarrow \overline{u}&=&F(u,\rho),                         \\
\label{der2}
where, &&
\left [\begin{array}{cc}
   a    &  b \\
    c     &  d
\end{array}  \right] \in SL(2,R), \ \frac{\partial F}{\partial u}>0.
\end{eqnarray}
\end{prop}

$F$  is appropriately smooth on the whole of
$\scri^{+} \simeq R \times S^{1}.$
$F$ is monotonic increasing in u for
each $\rho$, mapping the entire range for $u$ for
each $\rho-$generator to itself, and with non$-$vanishing
$u-$derivative so that the inverse transformation is also smooth.
The condition $\frac{\partial F}{\partial u}>0$ ensures that  the transformation
(\ref{der1}) preserves the orientation on each generator of $\scri^{+}$. The
group of transformations (\ref{der}), (\ref{der1}), (\ref{der2})
is the Newman$-$Unti (NU) group in 3 space$-$time dimensions.
The analogue of the NU group in 4 space$-$time dimensions has been defined in \cite{Pen2}, \cite{Pen3}, \cite{NU}.

$\mathcal C_{\scri^{+}}$ is determined,  via (\ref{con1}), by $\mathcal C_{
Min_{3}}$ and that is certainly preserved by $P(2,1)$, the Poincare group in  3 space$-$time dimensions.
Thus $P(2,1)$ is a subgroup of the NU group.



\subsection{
Self$-$diffeomorphisms of $\scri^{+}$ which preserve its strong conformal geometry
}
\label{selfdif}

Following Penrose \cite{Pen2}  it is possible to assign more geometric structure to $\scri^{+}$ than we have given hitherto. The preservation of this additional
structure will reduce the freedom in $F(u,\rho)$ in (\ref{der1}) to that of a function of $\alpha(\rho)$ only. This will give us the BMS group $  B(2,1).$

To this end it is natural to seek for quantities which
are invariants of the class $\mathcal C_{\scri^{+}}$, i.e., are the same
for
all members of the class $\mathcal C_{\scri^{+}}.$
In fact we can easily prove the following
\begin{prop}
The ratio $\frac{du}{dl}$ is independent of the choice of $\Omega$.
\end{prop}
The infinitesimal increments  $du$ and $dl$
were defined respectively in Lemma \ref{lemma2} and in (\ref{inmetr}), (\ref{ejisvsh}).

We can use this invariance to define the concept of ``null angle''  $\nu$  \cite{Pen2} between two
tangent directions at a point $P$ of $\scri^{+}$,
whose span contains the null normal direction to $\scri^{+}$,
by
\be
\nu=\frac{du}{dl}.
\ee
The induced degenerate metric (\ref{ejisvsh}) supplemented by the concept of null angle, provides the ``strong
conformal geometry''  \cite{Pen2}  at $\scri^{+}$.

We seek to find the group of self$-$diffeomorphisms of $\scri^{+}$ which preserves the
``strong conformal geometry'' at $\scri^{+}$, i.e., the group of diffeomorphisms of $\scri^{+}$
to itself which preserves angles and leaves invariant the ratio $\frac{du}{dl}.$

The allowed diffeomorphisms must have the form (\ref{der}), (\ref{der1}), (\ref{der2}) since these are the ones
preserving the inner conformal metric but the function $F$ must now have the special form that allows the ratio
$\frac{du}{dl}$  to remain invariant.

It is an easy exercise to prove that under (\ref{der}), (\ref{der1})
\be
\label{confac}
dl \longrightarrow K dl, \quad  where \normalfont \quad     K = \frac{ 1+ \rho^{\rm 2}}{     ( a \rho +  b)^{\rm 2}   +    ( c \rho +  d)^{\rm 2}                   }.
\ee
We must therefore also have
\be
du \longrightarrow  K du.
\ee
Since $K$ is independent of $u$, this means $u$ transforms according to
\be
u \longrightarrow  K  (u+\alpha(\rho)),
\ee
where $\alpha(\rho)$ is some real$-$valued function defined on $S^{1}$.

Let $\mathcal S(S^{1})$  be the space of real$-$valued functions defined on $S^{1}.$
The Theorem which follows
summarizes the previous results.

\begin{thrm}
\label{Theorem4}
The group of self$-$diffeomorphisms of $\scri^{+}$ which preserves the
``strong conformal geometry'' at $\scri^{+}$
is given by
\begin{eqnarray}
\label{der3}
\rho \longrightarrow \overline{\rho}&=&\frac{a \rho+b}{c \rho+d},  \\
\label{der4}
u \longrightarrow \overline{u}&=&\frac{ 1+ \rho^{\rm 2}}{     ( a \rho +  b)^{\rm 2}   +    ( c \rho +  d)^{\rm 2} }  (u+\alpha(\rho)),                         \\
\label{der5}
where, &&
\left [\begin{array}{cc}
   a    &  b \\
    c     &  d
\end{array}  \right] \in SL(2,R), \     \alpha(\rho)  \in \mathcal S(S^{1}).
\end{eqnarray}
\end{thrm}


Two remarks are in order regarding Theorem \ref{Theorem4}:
\begin{enumerate}

\item{Historically, the analogue of the classical action (\ref{der3}), (\ref{der4})
on  $\scri^{+}$  in the case of the original BMS group $B$  was given in \cite{Sachs1}
Equation (3.12), \cite{Sachs2} Equation (II.23), \cite{Pen1}  Equation (3.14).
}

\item{Since truly arbitrary supertranslation functions $\alpha(\rho)$ in (\ref{der4})
describe symmetry transformations in Penrose's sense, supertranslations
need not have some minimum degree of smoothness.}

\end{enumerate}


Let $\mathcal B$ be the set of
pairs $(\alpha(\rho),g)$, $\alpha(\rho)   \in  \mathcal S(S^{1})$,
$g=\left [\begin{array}{cc}
   a    &  b \\
    c     &  d
\end{array}  \right] \in SL(2,R).$
Application of two consecutive self$-$diffeomorphisms
\begin{equation}
\label{contr}
\left\{
    \begin{array}{l}
      \rho \longrightarrow \overline{\rho}\\
      u \longrightarrow \overline{u}
    \end{array}
  \right  \}, \quad
  \left\{
    \begin{array}{l}
      \overline{\rho} \longrightarrow   \overline{\overline{\rho}} \\
      \overline{u} \longrightarrow \overline{\overline{u}}
    \end{array}
  \right  \},
\end{equation}
defines an operation between the elements of $\mathcal B.$
It appears that $\mathcal B$ cannot be identified with $B(2,1)$ since
the action $T_{\mathcal B}$, resulting from the two consecutive self$-$diffeomorphisms
(\ref{contr}),
which defines the operation between the elements of $\mathcal B$, is not
identical to the action  $T$,
given by (\ref{actb}), (\ref{kfac}),
which defines
$B(2,1)$.

In order to bring home the identity of $\mathcal B$ with
$B(2,1)$ a convenient reformulation of
(\ref{der3}), (\ref{der4}) is needed.
This is given in section \ref{classical}.
Not surprisingly,
the reformulation needed,
is the analogue of
Equation (50)
in \cite{Bondi},
where $B$ firstly appeared as
a pseudo$-$group of local diffeomorphisms.



\subsection{``Null angles''
}

Let $P$ be  a point of $\scri^{+}$  and consider two non$-$null (spacelike)
tangent oriented  directions $ \gamma $ and $ \delta $ to $\scri^{+}$ at $P$.
Let
\begin{equation}
\label{normal}
N_{\alpha} = - \nabla_{\alpha} \Omega
\end{equation}
\it near \normalfont
$\scri^{+}.$ The vector field $N_{\alpha}$ is normal to the null hypersurface $\scri^{+}$.
Because $N_{\alpha}$ is orthogonal to itself ($N^{\alpha} N_{\alpha}=0   $), this vector field is also tangent to $\scri^{+}.$ The sign in
(\ref{normal}) is chosen so that $N^{\alpha}$ is future$-$directed
when $\Omega$ increases toward the future.

In 4 space$-$time dimensions there are two types of angle \cite{Pen3}  that  arise between $ \gamma $ and $ \delta. $ The first type occurs when the plane
defined by $ \gamma $ and $ \delta$  does not contain $N$. The angle between $ \gamma $ and $ \delta$
is measured with the intrinsic degenerate metric
on $\scri^{+}.$   The second type arises when the plane
defined by $ \gamma $ and $ \delta$   contains $N$ (neither of $ \gamma $ and $ \delta$ being actually in the direction of $N$). One can easily prove
that in this case
the angle between $ \gamma $ and $ \delta$
 measured with the intrinsic degenerate metric on $\scri^{+}$
 is always zero.

Although the angle between $ \gamma $ and $ \delta$ in this case is always zero, there is
still an invariant concept of separation between $ \gamma $ and $ \delta$.
This concept is precisely the concept of  ``null angle''.
$B$ is the group of self$-$diffeomorphisms of $\scri^{+}$
which leaves conformally invariant the intrinsic degenerate metric on $\scri^{+}$
and the ``null angle'' between any two  non$-$null
tangent directions at a point P of  $\scri^{+}$ which are
coplanar with
$N.$


In three space$-$time dimensions the situation is different.
One can prove the Proposition which follows.
\begin{prop}
In three space$-$time dimensions there is only one type of angle   that  arises between $ \gamma $ and $ \delta.$
The plane
spanned  by $ \gamma $ and $ \delta$  always  contains $N$. The angle between $ \gamma $ and $ \delta$
 measured with the intrinsic degenerate metric (\ref{ejisvsh}) on $\scri^{+}$
 is always zero. The ``null angle''
 $$
 \nu=\frac{du}{dl}
 $$
 gives an invariant concept of separation between $ \gamma $ and $ \delta$.
\end{prop}

\begin{prf}
The proof can be given with standard Linear Algebra and Differential Geometry and as such it is omitted.
\end{prf}

$B(2,1)$ is the group of self$-$diffeomorphisms of $\scri^{+}$
which leaves conformally invariant the intrinsic degenerate metric (\ref{ejisvsh}) on $\scri^{+}$
and the ``null angle'' between any two  non$-$null
tangent directions $ \gamma $ and $ \delta$ at a point P of  $\scri^{+}$.




\section{The classical action  $B(2,1) \times \scri^{+} \longrightarrow  \scri^{+}   $ }
\label{classical}

In order to
agree with notations used elsewhere (e.g. \cite{Pen2}, \cite{Pen3}) for the classical action
$B \times \scri^{+} \longrightarrow  \scri^{+}$ from the left, it now becomes convenient to use the action of
$G$ on  $\scri$
from the left.
Now,
we use (\ref{eqleft1}),     (\ref{eqleft2}), and
\be
r^{2}(\sigma)=\sigma_{1}^{2}+\sigma_{2}^{2}, \qquad r^{2}(g\sigma)=(a \sigma_{1} +  b  \sigma_{2})^{2}+(c \sigma_{1} +  d  \sigma_{2})^{2},
\ee
and we redefine
\be
\label{kfactor}
\kappa_{g}(\rho)= \frac{r^{2}(g\sigma)}{r^{2}(\sigma)}=\frac{(\rho a + b)^{2}+(\rho c + d)^{2}}{\rho^{2}+1}.
\ee

With these new definitions adapted to the left action of $G$  on $\scri$ one can now redefine
\begin{eqnarray}
\label{group}
B(2,1)&=&\mathcal H (P_{1}(R)) \circledS _{T} G,  \\
\label{actionnew}
(T(g) \alpha )(\rho) &=& \kappa_{g^{-1}}(\rho) \alpha (g^{-1}\rho).
\end{eqnarray}
$\scri^{+}$ can be identified with the product $R \times P_{1}(R).$ Let $[u,\rho]  \in  R \times P_{1}(R)=\scri^{+}.$
Then $(\alpha,g)  \in  B(2,1)$ may be obtained as the product $(\alpha,g)=(\alpha,Id)(0,g)$ (pure Lorentz followed by pure supertranslations).
The separate actions are, in order,
\begin{eqnarray}
(0,g)[u,\rho]&=&\left [ \kappa_{g}^{-1}(\rho)u, g\rho \right ] \in \scri^{+},  \\
(\alpha,Id)[u,\rho]&=&\left [ u+\alpha(\rho), \rho \right ] \in \scri^{+}.
\end{eqnarray}
Combining the two gives
\be
\label{combining}
(\alpha,g) \left [  u, \rho   \right ] = \left [  \kappa_{g}^{-1}(\rho)u + \alpha(g \rho), g \rho \right ].
\ee
This last formula, then gives the classical action $B(2,1) \times \scri^{+} \longrightarrow  \scri^{+}.$
Applying a further $(\beta,h)$ gives
\be
(\beta,h)(\alpha,g)=\left (\beta+T(h)\alpha,hg \right ),
\ee
where $T(h)\alpha$ is easily checked to be given by the action (\ref{actionnew}) defining $B(2,1)$ in (\ref{group}).
We note that $K=\kappa_{g}^{-1}(\rho)$, $K$, $\kappa_{g}(\rho)$ are given respectively by (\ref{confac}) and (\ref{kfactor}).
Moreover, (\ref{combining}) is a convenient reformulation of (\ref{der3}), (\ref{der4}).
An easy calculation shows
\begin{equation}
\label{combining1}
(T(g)\alpha,g) \left [  u, \rho   \right ] = \left [  \kappa_{g}^{-1}(\rho)(u +  \alpha( \rho)), g \rho \right ].
\end{equation}
The right$-$hand side of (\ref{combining1}) is identical to (\ref{der3}), (\ref{der4}).
We turn now to the study of $B(2,1).$

\section{The supertranslation subgroup}


\label{subgroup}

\indent

For notational convenience, hereafter,
we use the symbol $\mathcal A$ for
$\mathcal H (P_{1}(R)).$
As is well known, the set of functions
\begin{equation}
\label{basis}
e_{0}=\frac{1}{\sqrt{2 \pi}},  e_{n}= \frac{1}{\sqrt{ \pi} } \cos n\theta,  \varepsilon_{n}=\frac{1}{\sqrt{ \pi} } \sin n \theta, \ \ n=1,2,...,
\end{equation}
is a complete orthonormal basis for the Hilbert space $\mathcal A$.
Therefore, any function
$\alpha \in \mathcal A$ may be written as
\be
\alpha(\theta)=\sum_{n=0}^{\infty} (  c_{n} e_{n} + s_{n} \varepsilon_{n} ).
\ee

The inner product space $\mathcal A$ is an orthogonal direct sum of the subspaces $\mathcal T$ and $\mathcal S$,
the translation and proper
supertranslation subspaces respectively,
\be
\mathcal A = \mathcal T \bigoplus \mathcal S,
\ee
these subspaces consisting of the following functions:
\be
\mathcal T: t(\theta) = \sum_{n=0}^{1} (  c_{n} e_{n} + s_{n} \varepsilon_{n} )=a_{0} + a_{1}\cos \theta + a_{2}\sin \theta,
\ee
where the real numbers $a_{0},a_{1},a_{2}$ may be easily found from $c_{0},c_{1},s_{1}$ by using the formulae for the
$e_{0},e_{1},\varepsilon_{1}$,
\be
\mathcal S: s(\theta) = \sum_{n=2}^{\infty} (  c_{n} e_{n} + s_{n} \varepsilon_{n} ).
\ee

Therefore, every element $\alpha \in \mathcal A$ is uniquely 
decomposed as
\be
\label{dec}
\alpha=t+s
\ee
where $t \in \mathcal T$ and $s \in \mathcal S$. The elements $t$ and $s$ are the
orthogonal
projections of $\alpha$ on the spaces $\mathcal T$
and  $\mathcal S$ respectively:
\be
t=P_{\mathcal T}(\alpha), \quad t=P_{\mathcal S}(\alpha).
\ee
\newline
\noindent
$\mathcal T$ is invariant under the action  $T$ of $G$ on $\mathcal A$, given by
(\ref{actb}), whereas $\mathcal S$ is not.
Thus the orthogonal direct sum decomposition $\mathcal A = \mathcal T \bigoplus \mathcal S$ is not preserved by the $G$ action.
Restricting $\mathcal A$ to the $G$ invariant subspace $\mathcal T$ restricts $B(2,1)=\mathcal A
\circledS _{T} G$
to  $P(2,1)=\mathcal T \circledS _{T} G$, the double cover of the Poincare group in three space$-$time dimensions.
In fact the 3$-$tuplet $(a_{0},a_{1},a_{2})$ in the formula for $t(\theta) \in \mathcal T$  is the translation vector of
$P(2,1)$, and one quickly checks that the action $T$ of $G$ on these vectors gives the natural action of
$G/Z_{2}\approx L_{+}^{\uparrow}$, $L_{+}^{\uparrow}$ is the connected component of the identity of the
Lorentz group in three dimensions, on three$-$dimensional Minkowski space vectors.

\section{On the subgroups of $B(2,1)$
}
\label{subgroups}

\noindent


The following Proposition clarifies the relation of $B(2,1)$ with $\mathcal T$  and $ \mathcal A$.

\begin{prop}
\label{prop1}

\indent

The group B \rm{(2},\rm{1)} \normalfont \it has two normal subgroups and both are abelian:
\begin{enumerate}
\item $\mathcal T$ which is its unique 3$-$dimensional normal subgroup, and,
\item $\mathcal A$ which is infinite$-$dimensional.
\end{enumerate}
\end{prop}

\begin{prf}
The proof is similar to the one given by Sachs \cite{Sachs2} for the original BMS group $B$ and as such it is omitted.
\end{prf}

Property 1 identifies  $\mathcal T$  uniquely.
Obviously, the factor group $B(2,1)/\mathcal{A}$ is isomorphic to $G.$
$B(2,1)$ has an infinite$-$parameter family of Poincare subgroups, all of them
on equal footing. Not even the restricted Poincare group
$P(2,1)$ itself arises naturally as a subgroup of $B(2,1)$
in general. It is only in Minkowski space $\mathcal M$
that $P(2,1)$ is naturally singled out. In the case of
$\mathcal M$ we have some additional structure at
future null infinity $\mathfrak{J}^{+}$, namely the notion
of which cuts are to be labelled as ``good cuts''.
These ``good cuts'', which form a three$-$parameter family,
are obtained from one another by translations; but a
proper supertranslation
takes a ``good cut'' into
a ``bad cut''.
For such a supertranslation $ s \in \mathcal S, $
the six$-$parameter subgroup of $B(2,1)$, $\mathcal P = s^{-1}  P(2,1)  s$,
will be distinct from $P(2,1)$, though isomorphic with it.
For a general $s$, the subgroups $P(2,1)$ and $\mathcal P$ have only
the group of translations $\mathcal T$ in common.

In the case of the original BMS group $B$,
the infinite$-$parameter family of Poincare subgroups $P$
contained in $B$ \cite{Penrose}
underlies a new resolution \cite{haw} of the  black hole information paradox.
In 2+1$-$dimensional gravity with zero cosmological constant no black holes with
event horizons exist \cite{ABBOT}. However, as in the four$-$dimensional case,
the infinite$-$parameter family of  Poincare subgroups contained in $B \rm{(2},\rm{1)}$,
implies that in 2+1$-$dimensional quantum gravity with zero cosmological constant the vacuum is not unique.
In fact there is an infinite$-$parameter family of inequivalent vacua
which have zero energy but different angular momenta. None of these vacua are
preferred, and each   is annihilated by a different Poincare subgroup of $B \rm{(2},\rm{1)}.$

$\mathcal T$  is a normal subgroup of $\mathcal A$.
The quotient group
$PST=\mathcal A/ \mathcal T$
is
naturally isomorphic to  $\mathcal S$, the isomorphism being given by
\begin{equation}
\label{isom}
i: PST \rightarrow   \mathcal S, \quad
i(\alpha+\mathcal T)=  P_{\mathcal S}(\alpha),
\end{equation}
where $\alpha \in \mathcal A$, and $P_{\mathcal S}(\alpha)$ is the projection  of $\alpha$  onto $\mathcal S$.
$PST$ has as elements the proper supertranslations.
Isomorphism (\ref{isom})
identifies the infinite$-$dim$-$
ensional group
\be
\label{com}
PST(2,1)=PST \circledS _{T} G,
\ee
as a subgroup of $B\rm{(}\rm{2},\rm{1}\rm{)}$, with semidirect product specified by
\be
\label{actbnew}
T(g)(\alpha + \mathcal T)=T(g)\alpha +  \mathcal T,
\ee
where $\alpha \in \mathcal A$.

An immediate consequence of the definition (\ref{com}), (\ref{actbnew}) is the Corollary which follows.
Normal subgroup  and group isomorphism are denoted respectively by
$\trianglelefteq$ and $\simeq$.

\begin{cor}
\label{prop3}
Regarding the  group $PST(2,1)$ we have the following
\begin{enumerate}
\item $PST(2,1)\ntrianglelefteq B\rm{(}\rm{2},\rm{1}\rm{)}$
\item $PST \unlhd PST(2,1)$ \normalfont
\item $PST(2,1)/PST \simeq G$
\item $PST(2,1) \simeq B(2,1)/\mathcal T$
\end{enumerate}
\end{cor}

The IRS of $PST(2,1)$
describe quantum systems invariant under all
rigid translations in $R^{3}$.
Therefore, the IRS of $B$ are not associated with
elementary free particles. It seems plausible to suggest
that the carrier spaces of the IRS of $PST(2,1)$ have as elements
 quantum states of three-dimensional Chern$-$Simons gravity.
 In view of the relation $PST(2,1)/PST \simeq G$,
it is tempting
 to propose that $G$, which double covers $SO(2,1)$,
 contains within its structure a preferred way of representing
 a spin$-$like quantity which can be assigned to these
 quantum states.

\section{
The spaces $\mathcal A, \   \mathcal A^{\prime}, \  \hat{\mathcal A}  $
}
\label{spaces}

\noindent

A \it{character} \normalfont $\chi$ of $\mathcal A$ is a continuous homomorphism
\be
\chi: \mathcal A \rightarrow U(1)
\ee
of $\mathcal A$ into the multiplicative group of complex numbers of unit modulus. A composition law
\be
(\chi_{1} \chi_{2}) (\alpha) = \chi_{1}(\alpha) \chi_{2}(\alpha)
\ee
on the characters turns them into an abelian group $\hat{\mathcal A}$. Besides,
a multiplication by scalars
\be
(\kappa \chi)(\alpha)=\kappa \chi(\alpha), \ \kappa \in R,
\ee
turns the characters into a vector space over the reals.
By writing the character $\chi$ as
\be
\label{chi}
\chi(\alpha)=e^{i f(\alpha)},
\ee
one easily finds that the real$-$valued function $f$
1. Is continuous 2. Is additive , i.e., $f(\alpha_{1}+\alpha_{2})=f(\alpha_{1})+f(\alpha_{2})$, and, 3. It satisfies
$f(\kappa  \alpha) = \kappa f(\alpha), \ \kappa \in R.$
Therefore $f$ is a continuous linear functional on $\mathcal A$, that is, $f$ belongs to the topological dual $\mathcal A^{\prime}$ of $\mathcal A$. Thus the map
\begin{equation}
\label{isoj}
j: \mathcal A^{\prime}  \rightarrow  \hat{\mathcal A}, \ \
\it  (j(f))(\alpha)  =  e^{i f(\alpha)}, \
\end{equation}
$ \alpha \in \mathcal A,$ provides an isomorphism between $\hat{\mathcal A}$ and $\mathcal A^{\prime}$, as real vector spaces, and in particular as abelian groups.

From the Reisz$-$Fr$\acute{e}$chet theorem for Hilbert spaces, one concludes that \it{every} \normalfont element $f \in \mathcal A^{\prime}$ can be written uniquely in the form
\be
\label{ris}
f(\alpha)=<\beta,\alpha>,
\ee
where $\beta \in \mathcal A$ and $<\cdot,\cdot>$ denotes the
inner product of the Hilbert space $\mathcal A$. Thus the map
\begin{equation}
\label{isol}
l: \mathcal A  \rightarrow  \mathcal A^{\prime}, \ \
\it  (l(\beta))(\alpha)  =  <\beta,\alpha>,
\end{equation}
$\beta, \alpha \in \mathcal A$, provides an isomorphism between $\mathcal A$ and $\mathcal A^{\prime}$, as real vector spaces, and in particular as abelian groups.
Therefore, $\mathcal A^{\prime} \simeq
\mathcal H (P_{1}(R))
$.
Hereafter we will use the same symbol $\beta$ to denote
both an element of $\mathcal A$ and its associated
functional in $\mathcal A^{\prime}.$

We conclude that  
\begin{prop}
There are the isomorphisms
\begin{eqnarray}
\label{isos}
\mathcal A &  \! \! \! \underset{ l} {\longrightarrow}   \! \! \! & \mathcal A^{\prime}  \underset{ j} \longrightarrow  \hat{\mathcal A},
\end{eqnarray}
given by
$$
(l(\beta))(\alpha)  =  <\beta,\alpha>,
$$
and,
$$
(j(f))(\alpha)  =  e^{i f(\alpha)}.
$$

\end{prop}

\section{The dual action
$T^{\prime}$ and the supermomentum space $\mathcal A^{\prime}$}

\label{dualaction}

\indent

Requiring pointwise preservation of the isomorphism $\mathcal A^{\prime}  \underset{ j} \longrightarrow  \hat{\mathcal A}$
by the $G$ action determines uniquely a dual action $T^{\prime}$ of $G$ on $\mathcal A^{\prime}$.
We have the following
\begin{cor}
The dual action $T^{\prime}$ of $G$ on $\mathcal A^{\prime}$ is given by
\be
\label{dual}
<T^{\prime}(g) \beta, \alpha>= <\beta, T(g^{-1})\alpha>.
\ee
\end{cor}
\begin{prf}
The action $T$ of $G$ on $\mathcal A $ induces, in the usual way,
an action $\hat{T}$ of $G$ on $\hat{\mathcal A} $, defined by
\be
\label{dualnew}
(\hat{T} \chi)(\alpha)=\chi(T(g^{-1})\alpha).
\ee
From (\ref{isoj}) and (\ref{isol})  we infer that every character is of the form
\be
\chi(\alpha)=e^{i <\beta,\alpha>},
\ee
with $\beta$ uniquely determined
by the sequence of isomorphisms  $l$ and $j$
 given respectively by
 (\ref{isol}) and (\ref{isoj}).
Equation (\ref{dualnew}) implies that
the isomorphism
$\mathcal A^{\prime}  \underset{ j} \longrightarrow  \hat{\mathcal A}$ is preserved pointwise by the $G$ action, i.e.,
\be
(\hat{T}(g) \chi)(\alpha)=e^{i <T^{\prime}(g) \beta,\alpha>},
\ee
if and only if the action action $T^{\prime}$ of $G$ on $\mathcal A^{\prime}$ is given by
$$
<T^{\prime}(g) \beta, \alpha>= <\beta, T(g^{-1})\alpha>.
$$
This completes the proof.
\end{prf}

We readily find the dual action $T^{\prime}$ of $G$ on $\mathcal A^{\prime}$ as follows:
By using
(\ref{actb}) we obtain


\begin{eqnarray}
<T^{\prime}(g) \beta, \alpha>&=& <\beta, T(g^{-1})\alpha>=
\int
\beta(\rho)      (T(g^{-1})\alpha)(\rho)           \rm{d}\lambda(\rho) \nonumber \\
&&\int
\beta(\rho) \kappa_{g^{-1}}(\rho)\alpha(\rho g^{-1})
\rm{d}\lambda(\rho).
\end{eqnarray}
By using
(\ref{ratio}) we find
\be
<T^{\prime}(g) \beta, \alpha>=\int
\kappa_{g}^{-2}(\rho)    \beta(\rho g)        \alpha(\rho)
\rm{d}\lambda(\rho).
\ee
This holds for all $\alpha \in \mathcal A$, so we conclude
\be
\label{dual}
(T^{\prime}(g) \beta)(\rho)= \kappa_{g}^{-2}(\rho)    \beta(\rho g).
\ee
It is this dual action $T^{\prime}$ of $G$ on $\mathcal A^{\prime}$ which determines the structure of the IRS of
$B(2,1).$

$\mathcal A^{\prime}$ is isomorphic to
$\mathcal H (P_{1}(R)).$
This implies that the orthogonal  direct sum decomposition
$\mathcal A= \mathcal T \bigoplus \mathcal S$ dualizes to
\begin{equation}
\label{ordec}
\mathcal A^{\prime} = \mathcal T^{\prime} \bigoplus \mathcal S^{\prime},
\end{equation}
where $\mathcal T^{\prime}$ and $\mathcal S^{\prime}$ are the dual spaces of $\mathcal T$ and $\mathcal S$
respectively. Obviously we have
\begin{equation}
\mathcal T^{\prime}=\mathcal S^{\rm o}, \ \ \mathcal S^{\prime}=\mathcal T^{\rm o},
\end{equation}
where $\mathcal S^{\rm o}$
is the annihilator of $\mathcal S$, consisting of all linear functionals which vanish on $\mathcal S$.
$\mathcal T^{\rm o}$ is the annihilator of $\mathcal T.$ We have the isomorphisms
\begin{equation}
\mathcal A^{\prime}/ \mathcal T^{\prime} \approx \mathcal T^{\rm o} \ \ \rm{and} \ \ \normalfont \mathcal A^{\prime}/ \mathcal T^{\rm o}  \approx \mathcal T^{\prime}.
\end{equation}

It is of outmost significance that on passing
to the dual action $T^{\prime}$ of $G$ on $\mathcal A^{\prime}$, the subspace $\mathcal S^{\prime}$ becomes
invariant whereas the subspace $\mathcal T^{\prime} $
becomes \it noninvariant \normalfont (as it is easily infered from the dual action formula (\ref{dual}) had
the space $\mathcal S$ been invariant the space $\mathcal T^{\prime}$ would also have been invariant).
This explains in particular that there is no reason to expect that the $B(2,1)$ little groups should include
or be the same as  the $P(2,1)$ little groups: Since $\mathcal T^{\prime}$ is not  $T^{\prime}-$invariant
the usual action of $G$ on Poincare momentum space is not ``contained'' in the action of $G$ in $\mathcal A^{\prime}$.
In fact, as we will prove in Section \ref{compact}, all the little groups of $B(2,1)$  are  compact, whereas
the little groups of $P(2,1)$ are not all compact.

By using the isomorphism $\mathcal A^{\prime} \simeq \mathcal A = \mathcal H (P_{1}(R)),$
elements $\beta$ of $\mathcal A^{\prime}$
can be expanded, as the elements of $\mathcal A$ do, in the complete orthonormal basis (\ref{basis}):
\begin{equation}
\label{expansion}
 \beta(\theta) =
 p_{0} + p_{1}\cos \theta + p_{2}\sin \theta + \sum_{n=2}^{\infty} (  p_{n} e_{n} + \rm p_{n} \varepsilon_{n} ),
\end{equation}
where the decomposition $\mathcal A^{\prime} = \mathcal T^{\prime} \bigoplus \mathcal S^{\prime}$ has already been indicated:
The set $\{1, \cos \theta,\sin \theta \}$ is a spanning set of $\mathcal T^{\prime} $
and the set   $\{ e_{n}, \varepsilon_{n}, \ n=2,3,... \}$ is a spanning set of $\mathcal S^{\prime}.$
The isomorphism $ \mathcal A^{\prime}/ \mathcal T^{\rm o}  \approx \mathcal T^{\prime}$  suggests
that the triplet $(p_{0}, p_{1}, p_{2})$ may be thought of as the components of a ``momentum'' vector
with respect to the Poincare group $P(2,1)$. As a complement to this observation we note
that given that  $p=(p_{0}, p_{1}, p_{2})$ is a ``momentum'' vector then the character
$\chi(\alpha)=e^{i <p,\alpha>}$ associated to $p$ becomes a plane wave as the wave function of
a free particle with a well defined  momentum $p$ in the momentum representation we expect to be.

\section{Representation theory and its extension}
\indent

\label{representation}

\indent





Our construction of the IRS of $B(2,1)$ is based on G. W. Mackey's pioneering work on group representations
(see, for example, \cite{Mackey1},  \cite{Simms}, \cite{Isham},  \cite{Mackey2}, \cite{Mackey3}   ); in particular, is based on an extension to the relevant infinite-dimensional case
\cite{mac5}, \cite{Mel6}, \cite{Mel5}, \cite{McCMel} of his semidirect product theory.
It is interesting that the main sources of Mackey's  work were his extension  of the Stone$-$von Neumann theorem \cite{Stone}, which is the
cornerstone of quantum mechanics, Von Neuman's theory of direct integral decompositions
of operator algebras which he developed in the 1930s,
and Wigner's classic paper \cite{Wigner} in which he broke new grounds in mathematics by giving the first explicit treatment of infinite dimensional
representations of a Lie group.

\subsection{Wigner$-$Mackey's theory}
\label{inducing}

\indent

Let $\; \mathcal A \;$ and $\; \mathcal G \;$ be topological groups, and
let $\; T \;$ be a given homomorphism from $\; \mathcal G \;$ into
the group of automorphisms $\; {\rm Aut} (\mathcal A) \;$ of $\; \mathcal A. $
Suppose $\; \mathcal A \;$ is abelian and $\; \mathcal B= \mathcal A
\rtimes_{T} \mathcal G \;$ is the
semi$-$direct product of $\; \mathcal A \;$ and $\; \mathcal G, $ specified
by the continuous action $\; T: \mathcal G \longrightarrow
{\rm Aut} (\mathcal A). \;$ In the product topology of $\; \mathcal A \times
\mathcal G \;$, $\; \mathcal B \;$ then becomes a topological
group. It is assumed that it becomes a separable locally compact
topological group.

The action $T$ of
$ \mathcal G $ on $\mathcal A$ induces a dual action $T^{\prime}$ of $ \mathcal G  $ on $\mathcal A^{\prime}$ given by
(\ref{dual})
$$
<T^{\prime}(g) \beta, \alpha>= <\beta, T(g^{-1})\alpha>.
$$
For a  given functional
$ \beta \in A^{\prime},$ the largest subgroup $L_{\beta} $
of $ \mathcal  G $ which leaves $\beta $ fixed is called
the ``little group'' of $ \beta $, and the set of functionals which can be
reached from $ \beta $ by the
$ \mathcal  G $ action is called the orbit
of $ \beta $, denoted $ \mathcal G \beta.$


Let $U$ be a unitary representation of the ``little group''
$L_{\beta}$ on a Hilbert space $D.$
Let $ D_{\mu} $ be the space of functions $ \psi:
 \mathcal G \rightarrow D $ which satisfy the conditions
\begin{eqnarray}
(a) & \psi(gl)=U(l^{-1})\psi(g), & g \in  \mathcal G, \ l \in
L_{\beta}, \nonumber \\
& & \nonumber \\ (b) & \int_{ \mathcal G \beta}
<\psi(q),\psi(q)>{\rm d}\mu(q) < \infty,  \nonumber
\end{eqnarray}
where the scalar product under the integral sign is that of $D.$ Note, that the constraint $ (a) $ implies that $
<\psi(gl),\psi(gl)>=<\psi(g),\psi(g)>, $ and therefore the inner
product $<\psi(g),\psi(g)>,$ $ g \in  \mathcal G, $ is constant
along  every element $ q $ of the coset space $
\mathcal G/L_{\beta} \approx  \mathcal G \beta.$ This allows to assign a
meaning to $ <\psi(q),\psi(q)>, $  where $ q=gL_{\beta}, $ by
defining $ <\psi(q),\psi(q)>:= <\psi(g),\psi(g)>. $ Thus the
integrand in $ (b) $ becomes meaningful due to the condition
$ (a). $

A pre$-$Hilbert space structure can now be given to $\;
D_{\mu} \;$ by defining the scalar product
\begin{equation}
\label{hujkiolokijuhy} <\psi_{1},\psi_{2}>=\int_{ \mathcal G
\beta}<\psi_{1}(q),\psi_{2}(q)> {\rm d}\mu(q),
\end{equation}
where $\; \psi_{1},\psi_{2} \in D_{\mu}. \;$ It is convenient to
complete the space $\; D_{\mu} \;$ with respect to the norm
defined by the scalar product (\ref{hujkiolokijuhy}). In the
resulting Hilbert space, functions are identified whenever they
differ, at most, on a set of $\; \mu-\!$measure zero. Thus our
Hilbert space is
\begin{equation}
 D_{\mu}=L^{2}( \mathcal G \beta, \mu ,D).
\end{equation}

We note the following. Let $\mathcal T$ be  a
representation of $\mathcal B$
and let $Q$ and $V$ be its restriction to $\mathcal A$ and
$\mathcal G$ respectively, i.e.,
\be
Q(\alpha) \equiv \mathcal T(a), \quad V(g)\equiv \mathcal T(\rm{g}),
\ee
where, $a=(\alpha,I)$, $\rm{g} \normalfont = (\it{i},\it{g})$,
and $I$, $i$ are the identity elements of $\mathcal G$ and $\mathcal A$
respectively.
Since $(\alpha,g)=(\alpha,I)(i,g)$, for all $\alpha \in \mathcal A$ and $g \in \mathcal G,$ for any $h=(\alpha,g) \in \mathcal B$ we have
\be
\label{rep}
\mathcal T(h) \equiv  \mathcal T(\alpha,g)=Q(\alpha) V(g).
\ee

Thus a representation $\mathcal T$ of $\mathcal B$ is completely determined by its restriction $Q$ and $V$
to the subgroups $\mathcal A $ and $\mathcal G$ respectively.
Conversely representations $Q$ and $V$ of $\mathcal A$
and $\mathcal G$ correspondingly, define a unique representation
$\mathcal T$ of $\mathcal B$ by
(\ref{rep}). $\mathcal T$
is unitary when $Q$ and $V$ are unitary.

With this remark in mind we define now
actions of $\mathcal A$ and $\mathcal G$  on
$ D_{\mu} $ by
\begin{eqnarray}
\label{e1}
(Q (\alpha) \psi)(q) &=&  e^{i <g
\beta,\alpha> } \psi(q), \ \rm{and} \normalfont \\
\label{e2}
(V(g)
\psi)(q) & = & \sqrt{\frac{\rm{d}\mu_{\it{g}
}}{\rm{d}\mu}(q)} \psi(g
^{-1}q), \  \normalfont
\end{eqnarray}
respectively, where $ g
\in  \mathcal G, $ $q=g L_{\beta} \in \mathcal G/L_{\beta} \approx\mathcal G \beta,$ and
$\alpha \in \mathcal A$.
It is a straightforward exercise to verify that these actions
are in fact representations of $\mathcal A$ and $\mathcal G$ on
$ D_{\mu}.$
The ``Jacobian''
$\frac{\rm{d}\mu_{\it{g}
}
}{\rm{d}\mu} $
is known as the \it Radon$-$Nikodym \normalfont derivative of
$\mu_{g
}$ with respect to $\mu$ and ensures that the
representation $V$ is unitary.
Actions (\ref{e1}) and (\ref{e2})   define
a unitary representation of $\mathcal B$ on $D_{\mu}$
by
\begin{eqnarray}
(\mathcal T(h) \psi)(q) & = & (\mathcal T(\alpha,g) \psi)(q)=
((Q(\alpha) V(g)) \psi)(q)= (Q(\alpha) (V(g)\psi))(q) \nonumber \\
&=& e^{i <g
\beta,\alpha> } (V(g)\psi)(q) = \sqrt{\frac{\rm{d}\mu_{\it{g}
}}{\rm{d}\mu}(q)} e^{i <g
\beta,\alpha> }\psi(g
^{-1}q).
\end{eqnarray}


 The central results of induced representation theory
(\cite{Wigner}, \cite{Mackey1}, \cite{Simms},
\cite{Isham}, \cite{Mackey2}, \cite{Mackey3},
 and references therein) are the following:
\begin{enumerate}
\item {Given the topological restrictions on $ \mathcal B= \mathcal   A
\rtimes_{T}  \mathcal G $
(separability and local compactness), any representation $\mathcal T$ of $
\mathcal B, $  constructed by the method above,  is irreducible
if the representation $ U $ of $ L_{\beta} $ on $ D $
is irreducible. Thus an irreducible representation of $
\mathcal B $ is
induced
for {\it each} $ \beta \in
\mathcal A^{\prime} $
and {\it each} irreducible representation $ U $ of
$L_{\beta}.$} \item {If $ \mathcal B= \mathcal A
\rtimes_{T} \mathcal G $ is a
regular semi$-$direct product (i.e., if $
\mathcal A^{\prime} $ contains a Borel
subset which meets each $\mathcal G-$orbit in $
\mathcal A^{\prime} $
at
just one point) then {\it all} of its irreducible
representations
are obtained by this inducing construction.}
\end{enumerate}

It follows from the previous discussion that in order to give
the operators $\mathcal T$ of the IRS of $\mathcal B$  explicitly
it is
necessary to give the following information
\begin{enumerate}
\item{
An irreducible unitary representation $U$ of $L_{\beta}$ on a
Hilbert space $D$ for each ``little group'' $L_{\beta}.$}
\item{
A $  \mathcal G$$-$quasi$-$invariant measure $ \mu $
on each orbit $  \mathcal G \beta
\approx  \mathcal G / L_{\beta},$
where $L_{\beta} $ denotes the ``little group'' of the base point
$ \beta \in \mathcal A^{\prime} $ of the orbit
$ \mathcal G \beta.$
}
\end{enumerate}
In the case of
of
$ B(2,1)
$
then, in order to find its IRS $\mathcal T$, it is enough to provide the information cited in 1 and 2
for each of the orbit types.

\subsection{Extension of Wigner$-$Mackey's theory}
\label{extension}
\indent

The key feature of $B(2,1)$
is the enlargement of the three$-$dimensional vector space of rigid translations in $R^{2,1}$ to the infinite$-$dimensional
vector space of supertranslations
$ \mathcal H(P_{1}(R)) $. This turns
$B(2,1)$
into an infinite$-$di$-$ \newline    mensional Lie group.
In fact $\mathcal H (P_{1}(R))$ equipped with the standard
inner product (\ref{inpr})
is
a separable
\it infinite$-$dimensional \normalfont Hilbert space.
It is precisely this key feature which calls for an extension of Wigner$–$Mackey's theory.

Mackey assumes \cite{Mackey} that both
$\mathcal A$   and $\mathcal G$ in
$ \mathcal B= \mathcal A
\rtimes_{T} \mathcal G$ are locally compact.
In
the applications to Physics  it is assumed that $\mathcal A$ has an additional structure,
it is also a vector space, with vector addition being the group multiplication. It is well known that every Hausdorff topological vector space is locally compact if and only if
it is finite. This implies in particular that a Hausdorff infinite$-$dimensional topological vector space
is never locally compact.
In the case of $B(2,1)$
$\mathcal A$
is an infinite$-$dimensional  vector space.
In the whole class of topologies, physically motivated,
$\mathcal A$ can be  equipped with, $\mathcal A$
is Hausdorff and therefore not locally compact.
This renders Mackey's Theorems 14.1 and 14.2 in
\cite{Mackey} inapplicable.

In this paper $B(2,1)$ is equipped with the Hilbert topology. $\mathcal H (P_{1}(R))$
is endowed with the topology whose open sets are
the balls determined by the metric
induced by the scalar product (\ref{inpr}). Moreover, the group $G$, considered as a subset of $R^{4}$,
is endowed with the induced topology. In the product topology of $\mathcal H (P_{1}(R)) \times G$, the Hilbert topology,
$B(2,1)$ becomes a topological group; in particular it becomes
a non$-$locally compact group. The proof is similar to the
corresponding proof \cite{Cantoni} for the original BMS group $B$.

Since $B(2,1)$ is not locally compact in the Hilbert topology,
Mackey's Theorem
14.2 in \cite{Mackey} which proves
 irreducibility of the representations obtained by inducing
does not no longer applies. However, when  $B(2,1)$ is employed with the Hilbert topology we can prove that the representations of
$B(2,1)$ obtained by inducing
are irreducible notwithstanding the fact that $B(2,1)$ is not locally compact in the employed Hilbert topology. The proof
is practically the same with the corresponding proof given
in \cite{mac} for the original BMS group $B$ when $B$ is
endowed with the Hilbert topology.

Mackey proved, Theorem
14.1 in \cite{Mackey}, that only for regular semi$-$direct products $ \mathcal B= \mathcal A
\rtimes_{T} \mathcal G $ Wigner$-$Mackey inducing
 construction is exhaustive, i.e., that only for regular  $ \mathcal B$ \it  all \normalfont
 IRS of $ \mathcal B$ are obtained by inducing. Regular $\mathcal B$ means that  $\mathcal A^{\prime}$  can have no $\mathcal G$$-$quasi$-$invariant measures $\mu$ such that the action of $\mathcal G$ is strictly ergodic with
respect to $\mu$.
Whenever such measures $\mu$ do exist, an irreducible representation of $ \mathcal B$ may be associated
with each that is not obtained by inducing.
A sufficient condition which ensures the regularity of $ \mathcal B$  \cite{Mackey5}  (p. 43 and p. 141)    is the existence
of a Borel set in $\mathcal A^{\prime}$
that meets each  $\mathcal G$$-$orbit exactly once.

When the abelian normal subgroup $\mathcal A$ is also a
finite$-$dimensional vector space
finding a Borel
cross section which intersects each $\mathcal G-$orbit
in $\mathcal A^{\prime}$ exactly once is
fairly easy (e.g. \cite{Simms}).
However, when the abelian normal subgroup $\mathcal A$ is also an infinite$-$dimensional vector space, $\mathcal A^{\prime}$
is also an infinite$-$dimensional vector space,  it is not
generally feasible \cite{mac4}, \cite{Mel6} to classify completely the $\mathcal G-$orbits  in $\mathcal A^{\prime}$, let alone to search for a Borel set in $\mathcal A^{\prime}$ which meets each $\mathcal G-$orbit in $\mathcal A^{\prime}$ exactly once.

Consequently, the existence of a Borel
cross section which meets each $\mathcal G-$orbit in $\mathcal A^{\prime}$  exactly once cannot be used as a sufficient
condition which can be practically implemented when  $\mathcal A^{\prime}$ is infinite$-$dimensional
and new sufficient conditions which insure the exhaustivity
of the IRS of $\mathcal B$ constructed by Wigner$–$Mackey's ``little group'' method need to be introduced.

It is expected that it is the properties of the dual action of $\mathcal G$ on
$\mathcal A^{\prime}$ which will determine if the list of IRS obtained by inducing
is exhaustive. As a result we expect that the properties of the dual action of $\mathcal G$ on
$\mathcal A^{\prime}$
will imply  the validity of the sought new sufficient conditions.
In conclusion, when
$\mathcal A$ is
an
infinite$-$dimensional vector space,
proving
the exhaustivity of the IRS obtained by inducing
is reduced to the solution of the following two problems:
\begin{enumerate}
\item{Find sufficient conditions which ensure that $\mathcal A^{\prime}$  can have no $\mathcal G$$-$quasi$-$ \newline
invariant measures $\mu$ such that the action of $\mathcal G$ is strictly ergodic with
respect to $\mu$.}
\item{Find
those properties of the dual action of $\mathcal G$ on    $\mathcal A^{\prime}$  which imply the validity of the sufficient conditions stated in Problem 1.       }
\end{enumerate}

\noindent
Problems 1 and 2 were solved in two remarkable papers \cite{Piard1}, \cite{Piard2} by Piard, in the case where $\mathcal B$ is endowed
with the Hilbert topology.



\section{Topologies in $\mathcal A$}

\label{topologies}

\indent

The class of functions in $\mathcal A$ is not uniquely
determined. This allows  a wide range of ``reasonable'' topologies
for $\mathcal A.$

\subsection{$\mathcal A$ is not unique}

\indent

The key observation
is that Sachs's original derivation \cite{Sachs2}
 was superceded by
that of Penrose \cite{Pen1,Pen2,Pen3,pen2}.
As it was shown in subsection \ref{selfdif},
since
truly arbitrary supertranslation functions describe symmetry transformations in Penrose's sense,
\it supertranslations need not have some minimum degree of smoothness. \normalfont

Therefore, there is to a certain extent
an arbitrariness in the choice of the class of functions for $\mathcal A.$
Sachs's original derivation \cite{Sachs2} of the group $B$ required that the supertranslations be twice differentiable.
McCarthy widened \cite{mac3,mac4}
$C^{2}(S^{2},R)$
to $L^{2}(S^{2},R)$,
the square$-$integrable real$-$valued functions defined on the two sphere, in order to simplify
the treatment.

Other  choices for $\mathcal A$
are the function spaces $C^{k}(S^{2},R)$, $k=0,1,2,...$, \cite{Gi}, \cite{Cr}, the $k-$times
continuously differentiable functions on $S^{2}$, which are
physically plausible  in a gravitational
radiation context because they are adapted to
initial$-$value problems for hyperbolic differential equations, the
function space $C^{\infty}(S^{2},R)$ of smooth functions on $S^{2}$ \cite{mac5}, \cite{Gi}, \cite{Cr},
the function space $C^{\omega}(S^{2},R)$ of analytic functions on $S^{2}$ \cite{Mc}.




As it was
shown in section \ref{auto}
$B(2,1)$ can also be derived in Penrose's sense, i.e., as  that group
of \it exact \normalfont conformal motions of the future
(or past) null boundary $\scri^{+}$ (or $\scri^{-}$) of 3$-$dim
Minkowski
space$-$time
which is the automorphism group of the ``strong conformal geometry''
at $\scri^{+}$ (or $\scri^{-}$).
Consequently
similar remarks for the available freedom in the choice of
the degree of smoothness of the supertranslations apply also in
the case of $B(2,1).$ Thus besides $\mathcal H_{1}(P_{1}(R))$, other physically plausible
choices for $\mathcal A$ are the functions spaces $C^{k}(S^{1},R)$, $k=0,1,2,...$ ,
$C^{\infty}(S^{1},R)$, and $C^{\omega}(S^{1},R).$


\subsection{Topologies in $\mathcal A$}
\label{Top}

\indent

It is precisely the
arbitrariness in the class of functions allowed in $\mathcal A$ which permits a wide
range of choices of ``reasonable'' topologies for $\mathcal A$.
The only way to interpret the topologies is
to look at the structure of IRS
themselves.

The class of functions in $\mathcal A$ does not determine uniquely the
topology in $\mathcal A$. For the reasons set out in
\cite{mac3},    \cite{mac4}, \cite{mac5}, \cite{Gi}, \cite{Cr},
in the case of the original BMS group $B$
the Hilbert topology gives rise to IRS which are identified
with bound states, corresponding to elementary particles.
For the same reasons  the Hilbert topology given in $B(2,1)$ in this paper
leads to IRS which describe elementary particles.

If we choose for $\mathcal A$
the function spaces $C^{k}(S^{1},R)$, $k=0,1,2,...,$ as in the case of $B$ \cite{Cr},
\cite{Sob}, a physically reasonable class of topologies are the Sobolev topologies.
If we take $\mathcal A$ to be $C^{\infty}(S^{1},R)$, $\mathcal A^{\prime}$  is enlarged to the space of
generalised functions $\mathcal F(S^{1})$ on $S^{1}$,
and a natural choice,
as in the case of $B$ \cite{mac5}, \cite{Gi}, \cite{Cr}, is the nuclear topology.
The expectation is \cite{Cr} that in the nuclear topology new IRS are going to arise
which describe scattering states. If we choose for $\mathcal A$
the function space
$C^{\omega}(S^{1},R)$  rather than
the function space
$C^{\infty}(S^{1},R)$,
then, with an appropriately fine topology, $\mathcal A^{\prime}$
becomes enlarged
to the space $\mathcal Y(S^{1})$     of real hyperfunctions on $S^{1}.$
$\mathcal Y(S^{1})$ is larger than $\mathcal F(S^{1})$.
It seems \cite{Mc} that the category of hyperfunctions
is more appropriate than that of distributions
for discussing $S$-matrix theory.

\section{``Little groups'' are compact}

\label{compact}

\noindent


Before giving the proof that the ``little groups'' of $B(2,1)$ are compact
in the Hilbert topology, we make some comments
on the dependence of the ``little groups'' on the
topology $B(2,1)$ comes equipped with.

\subsection{ ``Little groups''  depend on the topology }

\noindent


``Little groups'' might change when the topology of $\mathcal A$ is modified: The induced
representations are associated with the existence of invariant characters, i.e., of elements in $\mathcal A^{'}$
which
are left invariant by the action of some ``little group''.
``Little groups'' are then connected, via $\mathcal A^{'}$, to the topology of $\mathcal A$.
Since, e.g., a refinement
of the topology may broaden $\mathcal A^{'}$, new invariant elements with associated little groups may come
into existence.

In fact this is precisely what happens in the case of the original BMS group
$B
$.
In the Hilbert topology all little groups of $B$ are compact \cite{mac3}, \cite{mac4}.
The effect of passing from the Hilbert (coarser) topology
to the nuclear (finer) topology is to enlarge $\mathcal A^{\prime}$
from $L^{2}(S^{2},R)$  to the space of generalised functions $\mathcal F(S^{2})$ on $S^{2}$, i.e.,
the distributions $\mathcal D(S^{2})=\mathcal F(S^{2})-L^{2}(S^{2},R)$ as well as
``regular'' $L^{2}(S^{2},R)$ functions. Non$-$compact little groups
arise in the nuclear topology and
are associated with distributional supermomenta $\chi \in \mathcal D(S^{2}).$
We expect that the same will happen in the case of $B(2,1)$.
In this section we prove that all ``little groups'' of $B(2,1)$
are compact in the Hilbert topology. We expect that if we pass
from the Hilbert to the nuclear topology
non$-$compact little groups are going to arise
in the
nuclear topology which will leave invariant distributional
supermomenta $\varphi \in \mathcal D(S^{1}).$
We also expect that if we choose for $\mathcal A$ the function space
$C^{\omega}(S^{1},R)$ non$-$compact little groups are going to arise which leave invariant
hyperfunctional supermomernta.

\subsection{Proof}

\indent

Recall that the little group
$L_{\beta}$ of any
$\beta \in \mathcal A^{\prime}$ is the set

\be
\label{lg}
L_{\beta}= \{ g \in G| \ T^{\prime}(g)\beta=\beta \}.
\ee

\noindent
By continuity of the representation $T^{\prime}$, $L_{\beta}$ is always a closed subgroup of $G.$
Evidently, if $\beta=0,$ $L_{\beta}=G.$
Representations of $B(2,1)$ that are induced  from
$\beta=0,$ and $L_{\beta}=G,$  are unfaithful
on the \it{whole} \normalfont  supertranslation subgroup, and in particular describe quantum systems invariant under all rigid translations in $R^{2,1}$. Clearly they
have no physical significance in the context of
$B(2,1)$ which is concerned with isolated sources.
So, hereafter, we always assume that $\beta \neq 0.$

We now prove that all little groups of $B(2,1)$ are
in fact compact in the Hilbert topology.
The standard topology on $G$ is given by the metric
derived from the norm
\be
\label{norm}
|g|=\left(|a|^{2}+ |b|^{2} +|c|^{2}+|d|^{2} \right)^{1/2}, \qquad   g=\begin{bmatrix}
a & b \\
c & d
\end{bmatrix} \in G.
\ee
The following result will be useful.
\begin{thrm}
\label{theoremkfactor}
If $L \subset G$ is a non$-$compact subgroup of $G$,
and has a fixed point in $\mathcal A^{\prime}$, there is
an infinite sequence $g_{n}$, $n=1,2,3,...,$
$g_{n} \subset L$ such that, for large enough $n,$
the function $\kappa_{g_{n} }(\rho)$ is greater
than any given $M>0$ over a region $R_{1}$ of $S^{1}$,
and $0<\kappa_{g_{n} }(\rho)\leq M$ over a region
$R_{2}$ of $S^{1}$, where $S^{1}=R_{1}\cup R_{2}$
(disjoint union), and the measure $\lambda(R_{2})$
is not greater than any given $\varepsilon>0.$
\end{thrm}
\begin{prf}
The proof is similar to the corresponding proof for $B(2,2)$ \cite{McCMel} and as such it is omitted.
\end{prf}

The $\kappa$
factor corresponds physically to a
``redshift factor''. Theorem \ref{theoremkfactor} amounts
to the anticipated fact that a sequence of Lorentz transformations which contains a sequence of pure boosts
with velocity parameter tending arbitrary close to the
speed of light subjects a region of the one$-$dimensional
``selestial sphere'' with arbitrary small complement to arbitrary large ``blue$-$ \newline shifts'' (the complement is
``redshifted'').

We now prove the following:
\begin{thrm}
\label{com}
Every little group of $B(2,1)$ is compact.
\end{thrm}
\begin{prf} The proof proceeds by contradiction.
Let $\beta \in \mathcal A^{\prime}- \bf 0$ have little group
$L_{\beta}.$ It will be shown that if $L_{\beta}$ is non$-$compact
and $\beta$  is fixed under $L_{\beta}$, then $\beta =0$ or
$\beta$  cannot be square integrable.
A somewhat stronger result will, in fact,
be proved, namely that $\beta=0$ or $|\beta|$
cannot be integrable. This implies the desired
result since, from $(1-\beta)^{2}\geq 0,$ one
gets $|\beta|\leq \frac{1}{2} (1+ \beta^{2}).$

Assume, contrary to the theorem, that $L_{\beta}$ is non$-$compact for some $\beta \neq 0.$
Then $L_{\beta} \subset G$  must be unbounded with respect to the standard metric (\ref{norm}) for $G.$
Indeed, $L_{\beta} $ is closed, so if it were bounded, it would be compact, contrary to our
starting assumption.
By definition $(T^{\prime}(g) \beta)(\rho)=\beta(\rho)$
for all $g \in L_{\beta}.$ Hence, for all $g \in L_{\beta},$
\be
\int_{\it S^{\rm 1}
} |(T^{\prime}(g) \beta)(\rho)| \rm d \lambda(\rho)=\int_{\it S^{\rm 1}
} |\beta (\rho)|      \rm d \lambda(\rho)
\ee
for all $g \in L_{\beta}.$  Using  (\ref{dual}) to substitute for $T^{\prime}(g) \beta$, changing the variable $\rho \mapsto \rho g^{-1}$,
and using  (\ref{ratio}), gives
\be
\int_{\it S^{\rm 1}
}   \kappa_{g^{-1}}  (\rho)    |\beta (\rho)| \rm d \lambda(\rho)=\int_{\it S^{\rm 1}
} |\beta (\rho)|      \rm d \lambda(\rho)
\ee
for all $g \in L_{\beta}.$ Since $L_{\beta}$ is a group, the same equation applies with $g^{-1}$ replaced by $g$. Then the previous
equation becomes
\be
\int_{S^{1}
}  ( \kappa_{g}  (\rho) -1)   |\beta (\rho)| \rm d \lambda(\rho)=0
\ee
for all $g \in L_{\beta}.$ Then using Theorem \ref{theoremkfactor} for $\kappa_{g}  (\rho)$, the rest of the proof
is essentially the same as the corresponding proof for $B(2,2)$ in \cite{McCMel}. This completes the proof.


\end{prf}

Theorem \ref{com} intuitively can be understood intuitively as follows:
The dual action (\ref{dual}) $T^{\prime}$  of $G$  on  $ \mathcal H (P_{1}(R))$
is similar \cite{Mel6}  to   the transformation law  for light intensity
distributions on distant $1-dim$ ``celestial'' sphere. A non$-$compact
little group $L_{\beta}$ contains a sequence of boosts with velocity which becomes arbitrary close to the speed of light.
For an intensity distribution $\beta$ to be fixed under $L_{\beta}$ it must be such that when subjected to
arbitrary large ``blueshifts'' over a region of the $1-dim$ ``celestial'' sphere with arbitrary small complement
(the complement is ``redshifted''), it remains the same. It is not surprising that there are no such non$-$zero
$\beta$'s, so that there cannot be any non$-$compact little groups.
However, by the same reasoning, there may exist  \it distributional
intensity ditributions \normalfont $\beta$ invariant under non$-$compact little groups.This is
precisely what we expect to happen if we take $\mathcal A $ to be
$C^{\infty}(S^{1},R)$ and equip it with the nuclear topology.

\section{The ``Little groups'' of $B(2,1)$}

\label{littlegroups}

In section \ref{compact} it was shown that the ``little groups''
of $B(2,1)$   are compact subgroups of $\rm SL(2, \it R).$ The maximal compact subgroup of $G=\rm SL(2, \it R)$ is just the subgroup $SO(2).$ That is to say,
if $H$ is a compact subgroup of $G,$ then some conjugate  $gHg^{-1}, g \in G,$ of $H$ is a compact subgroup of $SO(2).$ In the representation theory of $B(2,1)$, the ``little groups'' are only significant up to conjugacy. So, after a possible conjugation, we may take every
``little group'' to be  a compact (or equivalently closed)
subgroup of $SO(2).$ Therefore, we need to find the compact subgroups of $SO(2).$

\subsection{The compact subgroups of $SO(2)$}

We prove now the following
\begin{prop}
\label{propsub}
The compact subgroups of $SO(2)$ are classified as follows:
\begin{enumerate}

\item{$SO(2)$ is the only one$-$dimensional compact subgroup.}

\item{$C_{n},$ $n \geq 1,$   the finite cyclic groups of order $n$, are the only zero$-$dimensional compact subgroups. }

\end{enumerate}
\end{prop}

\begin{prf}

$SO(2)$ is the set of matrices of the form
\begin{equation}
g(\omega) =   g=\begin{bmatrix}
\ \ \cos \omega & \sin \omega \\
- \sin \omega  & \cos \omega
\end{bmatrix} \in G.
\end{equation}
where $\omega$ is any real number, determined mod 2$\pi$, and may be taken in the range $0  \leq \omega \leq 2 \pi.  $
\end{prf}
The infinitesimal generator of $SO(2)$ is
\begin{equation} \left . \frac{\rm d \it g(\omega)}{\rm d \omega} \right |_{\omega=0}      =
\begin{bmatrix}
\ \ 0  & 1 \\
- 1  & 0
\end{bmatrix}=J.
\end{equation}

Hence the Lie algebra $\mathcal L K$ of $K=SO(2)$ is the vector space of all matrices of the form
\begin{equation}
\alpha J=
\begin{bmatrix}
\ \ 0  & \alpha \\
- \alpha  & 0
\end{bmatrix},
\end{equation}
where $\alpha$ is any real number. The exponential map $exp: \mathcal L K \longrightarrow K$
here is the usual exponential for matrices. This is given by
\begin{equation}
exp \left ( \alpha J \right ) =g(\alpha).
\end{equation}
This map is obviously onto but it is not one$-$to$-$one.

We now wish to describe the compact (equivalently closed) subgroups of $K.$
Every closed subgroup of a Lie group is a Lie subgroup (see e.g. \cite{Helgason}).
Let $H$ be any closed subgroup of $K$, $H \subset K$, with dimension 1.
The Lie algebra $\mathcal L H$ of $H$ has dimension 1, so must coincide with
$\mathcal L K$; $\mathcal L H=\mathcal L K$. $K$ is a connected topological group,
so \it any \normalfont neighborhood of the identity of $K$ generates the whole of
$K$ (see e.g. \cite{Hausner}). But $ exp \left (  \mathcal L H \right )= exp \left (  \mathcal L K \right )    $
contains a neighborhood of the identity of $H$, and also of $K$. The intersection of these is
a neighborhood of the identity of both groups. This intersection belongs to $H$, and generates the whole of
$K$. So  $K  \subset H$. But  $H  \subset K$. Therefore, $H=K.$

The only remaining case to consider is that in which $H$ is zero$-$dimensional.
Then $H$ is a discrete closed subgroup of the compact group $K$ and so, necessarily,
a finite subgroup $F \subset K$. But the only finite discrete subgroups of $K$ are the
finite cyclic groups $C_{n}$ of order   $n$,   $n \geq 1.$  Therefore $F=C_{n}$,  $n \geq 1.$
This completes the proof.

\subsection{The ``Little groups'' of $B(2,1)$}

In \cite{paper2}   we prove that not all the compact subgroups of $SO(2)$,
given in Proposition \ref{propsub}, qualify as little groups.
We prove in particular the following

\begin{thrm}
The ``little groups'' of $B(2,1)$ are
\begin{enumerate}
\item{SO(2) itself.}
\item{The cyclic groups $C_{n}$  of even order. }
\end{enumerate}
\end{thrm}

\section{Exhaustivity of Induced Representations}

\label{exhaustitivity}


\indent

Proving exhaustivity opens a whole new chapter in the theory of dynamical systems.
The dynamical system at hand is the triple $(\mathcal A^{\prime}, \mathcal M , \mathcal G)$ where $\mathcal M$ is a class
of
$\mathcal G-$quasi$-$invariant
measures on $\mathcal A^{\prime}$. It is the properties of the  $\mathcal G$ action on $\mathcal A^{\prime}$ which determine  if the
list of IRS of $ \mathcal B= \mathcal A \rtimes_{T} \mathcal G$  obtained by
inducing
is exhaustive.

Piard with a view to proving exhaustivity of McCarthy's list of representations \cite{mac3}, \cite{mac4},
studied class of semi$-$direct products of the form $ \mathcal B= \mathcal A \rtimes_{T} \mathcal G$,
where $\mathcal A$  is either $C^{\infty}(S^{1},R)$ endowed with the nuclear topology, or
$\mathcal A$  is the Hilbert space of a Gelfand triple
$ \mathcal A_{\rm o}  \subset  \mathcal A  \subset  \mathcal A_{\rm o}^{\prime} $
($ \mathcal A_{\rm o}$ is nuclear dense in $ \mathcal A$). $\mathcal A$
in the latter case is a separable
countable
\it infinite$-$dimensional \normalfont Hilbert space
endowed with the
strong topology which is also known in the literature
as metric or norm topology. $\mathcal G$ in both cases is assumed to be locally compact.
Both the original BMS group $B$ in the Hilbert topology and $B(2,1)$
in the Hilbert topology fall into this latter class of groups.


In a nutshell Piard associates a cylindrical measure
on $\mathcal A^{\prime}$ (respectively $\mathcal A_{\rm o}^{\prime}$  in which case the measure is concentrated on $\mathcal A_{\rm o}^{\prime} \cap \mathcal A^{\prime}$). When the representation of
$\mathcal B$ is irreducible this measure is ergodic
for the dual action of $\mathcal G$ on $\mathcal A^{\prime}$ and under the assumption that the measure
is concentrated on an orbit Piard proves that the
representation  of
$\mathcal B$ is induced. When $\mathcal A$ is a
Hilbert space Piard gives sufficient conditions
that all these ergodic measures be concentrated on an
orbit and hence that all IRS of $\mathcal B$ be induced.






In \cite{Piard2}
Piard by using
the properties of the $\mathcal G$
action in $\mathcal A^{\prime}$
proved that the
sufficient conditions introduced in \cite{Piard1} are verified by $B.$
He thus solved, in the case of $B$, problems 1 and 2 stated in subsection \ref{extension}.
In the case of $B$, $\mathcal G=SL(2,C)$ and
$\mathcal A^{\prime}\simeq L^{2}(S^{2},R).$
Therefore,
Piard concluded,
McCarthy's list of IRS in the Hilbert topology \cite{mac3}, \cite{mac4} is exhaustive. In this
section, by using the properties of the dual action of $SL(2,R)$
on $\mathcal A^{\prime} \simeq \mathcal H (P_{1}(R)),$
 we prove that the sufficient conditions introduced in \cite{Piard1} are also verified by $B(2,1).$

\subsection{Properties of the dual action}

\indent

Three are the properties of the dual action of $SL(2,R)$ on $\mathcal A^{\prime}$ which suffice to insure the
validity of the sufficient conditions introduced in \cite{Piard1}.  These properties are stated in the three Propositions which follow.
\begin{prop}
\label{propsub1}
Let $g_{n},n=1,2,3,...,$ be a sequence in $G$ such that $|g_{n}| \longrightarrow   \infty.$  Then there exists a subsequence $g_{n_{p}}$ such that $\kappa_{g_{n_{p}}}(\rho)\longrightarrow   \infty$ almost uniformly
as $p \longrightarrow \infty$.
\end{prop}

\begin{prf}
The proof is the same as the corresponding proof for the original BMS group $B$ given in \cite{Piard2}.
\end{prf}


Proposition \ref{propsub1}
is an immediate
consequence  of Theorem \ref{theoremkfactor}
and asserts that \it every \normalfont  sequence of Lorentz transformations which
subjects a region of the one$-$dimensional
``selestial sphere'' with arbitrary small complement to arbitrary large ``blueshifts''
contains a subsequence of Lorentz transformations which
subjects
a region of the one$-$dimensional
``selestial sphere''
with
complement of measure zero to infinite ``blueshift''.

\begin{prop}
\label{propsub2}
Let $g_{n},n=1,2,3,...,$ be a sequence in $G$ such that $|g_{n}| \longrightarrow   \infty.$  Then there exists a subsequence $g_{n_{p}}$
such that $| | T^{\prime}(  g_{n_{p}})\beta || \longrightarrow \infty$ almost uniformly
as $p \longrightarrow \infty$ for all $\beta \in \mathcal A^{\prime}.
$
\end{prop}
\begin{prf}
From (\ref{dual}) we obtain
\be
\label{length}
||(T^{\prime}(g_{n}) \beta||^{2}= \int_{S^{1}
} \kappa_{g_{n}}^{-4}(\rho) \beta^{2}(\rho g_{n})   d \lambda(\rho).
\ee
Using (\ref{ratio}) and changing  the variable $\rho \mapsto \xi=\rho g_{n}$ yields
\be
\label{length1}
||(T^{\prime}(g_{n}) \beta||^{2}= \int_{S^{1}
} \kappa_{g_{n}}^{-3}(\xi g^{-1}_{n} ) \beta^{2}(\xi)   d \lambda(\xi).
\ee
From (\ref{actb}) we have
\begin{eqnarray}
[(T(g^{-1}_{n}) T(g_{n}))\alpha](\xi) &=& [T(g^{-1}_{n}) (T(g_{n})\alpha)](\xi)=\kappa_{g^{-1}_{n}} (\xi)[T(g_{n})\alpha](\xi g^{-1}_{n} )
\nonumber \\
 &=& \kappa_{g^{-1}_{n}}(\xi)   \kappa_{g_{n}}  (\xi g^{-1}_{n} )             \alpha(\xi )
\end{eqnarray}
for all $\alpha \in \mathcal A.$ Therefore, $\kappa_{g^{-1}_{n}}(\xi)   \kappa_{g_{n}}  (\xi g^{-1}_{n} ) =1  $ almost everywhere. Henceforth we obtain
\be
\label{length2}
||(T^{\prime}(g_{n}) \beta||^{2}= \int_{S^{1}
} \kappa_{g^{-1}_{n}}^{3}(\xi  ) \beta^{2}(\xi)   d \lambda(\xi).
\ee
The rest of the proof is essentially the same as the corresponding proof for the original BMS group $B$ given in \cite{Piard2}.


\end{prf}

Proposition \ref{propsub2}
is a
consequence  of Proposition \ref{propsub1}
and states
that \it every \normalfont  sequence of Lorentz transformations which
subjects a region of the one$-$dimen$-$ \newline sional
``selestial sphere'' with arbitrary small complement to arbitrary large ``blueshifts''
contains a subsequence of Lorentz transformations which
transforms
\it any \normalfont square integrable  intensity  distribution
to a distribution of infinite intensity
in a region of the one$-$dimensional
``selestial sphere''
with
complement of measure zero.

A consequence of Proposition \ref{propsub2} which
refers to
\it any \normalfont
sequence of
square integrable  intensity  distributions
on the one$-$dimensional ``selestial sphere'' is
given in the Proposition which follows.


\begin{prop}
\label{propsub3}
Let $\beta_{n}$, $n=1,2,3,...,$   $\beta_{n}\in \mathcal A^{\prime},$                    be a sequence
weakly convergent to
$\beta_{0}$ and $g_{n}$, $n=1,2,3,...,$ a sequence in $G$ such that  $|g_{n}| \longrightarrow \infty$. Then there exists a subsequence $g_{n_{p}}$ such that
$||T^{\prime}(  g_{n_{p}}) \beta_{n_{p}}  || \longrightarrow \infty$ almost uniformly   as $p \longrightarrow \infty.$
\end{prop}

\begin{prf}
The proof is practically the same as the corresponding proof for the original BMS group $B$ given in \cite{Piard2}.
\end{prf}



\subsection{Sufficient conditions are satisfied}

\indent

Piard's sufficient conditions introduced in \cite{Piard1} are:
\begin{enumerate}

\item{The $\mathcal G-$orbits in $\mathcal A^{\prime}$  are open relatively to their strong closure.}

\item{For any two minimal invariant cones $C_{\beta_{1}}$ and $C_{\beta_{2}}$ such that $C_{\beta_{1}} \cap  C_{\beta_{1}}= \{ 0 \},           $ their intersections $F_{\beta_{1}}$ and $F_{\beta_{2}}$ with the unit ball are such that $      \bar{F}_{\beta_{1}} \neq  \bar{F}_{\beta_{2}}$ where $\bar{F}_{\beta_{i}} $   is the weak closure of $F_{\beta_{i}}$.}

\end{enumerate}

The minimal invariant cone $C_{\beta}$ is the smallest cone invariant
under the $\mathcal G-$action in $\mathcal A^{\prime}$ which contains $\beta:$
$$
C_{\beta}=\{  \lambda (T^{\prime}(g)\beta), \  g \in \mathcal G, \ \lambda \in  R^{+} \}.
$$
\noindent
$\mathcal A^{\prime}$ is equipped with the weak topology. In the case at hand $\mathcal G=SL(2,R)$,
$\mathcal A=\mathcal H (P_{1}(R))$, and $\mathcal A^{\prime}\simeq \mathcal H (P_{1}(R)).$

By using Propositions \ref{propsub2} and \ref{propsub3}  we arrive at the following
\begin{prop}
\label{propsub4}
Piard's sufficient conditions 1 and 2 are satisfied in the case of $B(2,1).$
\end{prop}

\begin{prf}
The proof is essentially the same as the corresponding proof for the original BMS group $B$ given in \cite{Piard2}.
\end{prf}

Consequently  all the cylindrical ergodic measures on    $\mathcal A^{\prime}$  associated to the IRS of  $B(2,1)$
are concentrated on the $SL(2,R)$$-$orbits on  $\mathcal A^{\prime}$. These measures correspond to subsets of
$\mathcal A^{\prime}$ which are ``indecomposable'' in the set theoretic sense.
We conclude  \cite{Piard1}, \cite{Piard2} that the following Theorem holds:

\begin{thrm}
All IRS of $B(2,1)$ are induced.
\end{thrm}

Thus \it all \normalfont IRS of $B(2,1)$ are obtained by the Wigner$-$Mackey's  inducing construction given in
subsection \ref{inducing}.




\normalfont

\section{Construction of the IRS of $B(2,1)$}

\label{construction}

\indent
To find explicitly the operators of the induced
representations of $B(2,1)$, it suffices to provide the information
cited in (i) and (ii) in subsection \ref{inducing} for each of the orbit types. We note that
all the little groups are abelian. All IRS of an abelian group are one$-$dimensional.
\newline

\noindent
\bf (i) \normalfont The IRS of the ``Little groups'', i.e., of
$SO(2)$ and of $C_{n},$ $n$  even, have been given in \cite{Mel6}.

$\newline$


\noindent
\bf (ii) \normalfont A $ \mathcal G$$-$quasi$-$invariant
measure $ \mu $ on each orbit $ \mathcal G \phi \approx
\mathcal G / L_{\phi}$ is required, however, a $ \mathcal G$$-$invariant
measure $ \mu $ on each orbit $ \mathcal G \phi \approx
\mathcal G / L_{\phi} $ will be given in all cases. We note that when $\mu$ is $ \mathcal G$$-$invariant then
\be
\label{mo}
\frac{\rm{d}\mu_{\it{g}
}}{\rm{d}\mu}(q)=1, \quad g \in \mathcal G, \quad   q  \in  \mathcal G \phi.
\ee
This is precisely what happens in the case of $B(2,1)$ where $\mathcal G = G = \rm SL(2, \it R),$
and $L_{\phi}$ is either $SO(2)$ or $C_{n},$ $n$  even.
(\ref{mo}) applies in particular in (\ref{e2}).
In \cite{Mel6} $G$$-$invariant
measures  on each orbit $  G \phi \approx
 G / L_{\phi}$ were obtained by projecting the unique (up to a multiplicative constant)
Haar measure of $G$ to the orbits
$ G \phi \approx G / L_{\phi}$
via the canonical map
$$
 G \longrightarrow G/L_{\phi}.
$$
However, explicit expressions for the $G$$-$invariant measures on the orbits
$ G / L_{\phi}$ were not given in \cite{Mel6}.
Here we give explicit expressions for these measures. Apart from their own interest,
these measures will be
particularly useful when we make contact with other approaches to quantum gravity,
e.g. with canonical quantum gravity \cite{can}.


$\newline$

\noindent
\bf (ii$_{\alpha}$) \normalfont
The orbit $01 \equiv G/L_{\phi}$, $L_{\phi}=K=SO(2)$, is the coset space
$\mathcal S={\rm S}{\rm L}(2,R)/SO(2)$. It is well known \cite{hel}
that every element
$\gamma$ of $\mathcal S$ can be uniquely written in the form
\begin{equation}
\gamma=n a (n a)^{T},
\end{equation}
where,
\begin{equation}
n \in N=
\left(
\begin{array}{cc}
1  & x \\
0  &  1
\end{array}
\right)\cong R, \quad a \in A=
\left(
\begin{array}{cc}
y  & 0 \\
0  &  \frac{1}{y}
\end{array}
\right)\cong R_{>0},
\end{equation}
$x \in R, \ y > 0,$ and $T$
denotes transposition.
$\mathcal S$ can be identified with the
Poincare half$-$plane
\begin{equation}
H= \{ (x,y)|y>0, \ x,y \in R \}
\end{equation}
via the map
\begin{equation}
\label{bij}
(x,y) \longrightarrow
\left(
\begin{array}{cc}
1  & x \\
0  &  1
\end{array}
\right) \left(
\begin{array}{cc}
y  & 0 \\
0  &  \frac{1}{y}
\end{array}
\right)
\left ( \left(
\begin{array}{cc}
1  & x \\
0  &  1
\end{array}
\right) \left(
\begin{array}{cc}
y  & 0 \\
0  &  \frac{1}{y}
\end{array}
\right)   \right )^{T}.
\end{equation}
A $G$$-$invariant measure on $H$ is given by
\begin{equation}
{\rm d}\mu= \frac{{\rm d}x \wedge {\rm
d}y }{y^{2}}.
\end{equation}
This is also, given  (\ref{bij}),  a $G$$-$invariant measure on $\mathcal S.$

$\newline$

\noindent
\bf (ii$_{\beta}$) \normalfont
A $ G$$-$invariant measure on
$$G={\rm S}{\rm L}(2,R)= \left \{ \left(
\begin{array}{cc}
a  & b \\
c   &  d
\end{array}
\right), \quad a,b,c,d \in R, \quad ad-bc=1 \right \}, \;$$
is given by \begin{equation}
\label{lyhgbhgblyuitg7lyugk} {\rm d}g= \frac{{\rm d}a \wedge {\rm
d}b \wedge {\rm d}c}{a}. \end{equation}

The orbits $02 \equiv G/L_{\phi}$,      $L_{\phi}=C_{n}$, where $n$ is even,  can be endowed with the $G$$-$invariant measure on $G$ given in
(\ref{lyhgbhgblyuitg7lyugk}).
 Indeed, for a given little group $L_{\phi}$,  the orbit
$02 \equiv G/L_{\phi}$ is the space of orbits of the right action
$T : G \times L_{\phi} \rightarrow G$
 of the group $L_{\phi}$ on $G$
given by
\begin{equation}
\label{action}
g \star c := g \cdot c,
\end{equation}
where $g \in G$, and $c \in L_{\phi}$.
Thus the action $T$ denoted by $\star$ is identical to the group multiplication in $G$. Since the group $ L_{\phi}$ is finite and since the action (\ref{action}) is fixed point free the coset spaces $G/L_{\phi}$ inherit the measure on $G$.\emph{}
\newline

\indent
This completes the necessary information in order to construct the induced representations of $B(2,1)$.
To conclude, in this paper \it all \normalfont IRS
of $B(2, 1)$ have been constructed in the Hilbert topology.



\section{Connections to Geometry}

\label{geometry}

\indent

In three space$-$time dimensions with zero cosmological constant
when the Einstein tensor vanishes so does the
four$-$index Riemann$-$Christoffel curvature tensor. As a result, all
three$-$dimensional Einstein spaces with zero cosmological constant  are flat.
Thus there are no local
gravitational degrees of freedom but a finite number of global degrees of freedom which arise
from the topology and matter content of the space$-$time.

In the case of the four$-$dimensional  asymptotically flat space$-$times
Foster showed \cite{Fo}, \cite{Fo1} that
when Penrose's future null infinity
$\scri^{+}$  is homeomorphic to $R \times S^{2}$ there exists an asymptotic symmetry group,
the original BMS group $B$, and vice versa. For other homeomorphic types of
$\scri^{+}$  there is no in general asymptotic symmetry group and when it exists
the asymptotic symmetry group
neither is it determined uniquely nor is it $B.$

By following through Foster's analysis
one can easily show that in the case of
three$-$dimensional   flat space$-$times a similar result holds:
$B(2,1)$ is the asymptotic symmetry group if and only if $\scri^{+}$  is homeomorphic to $R \times S^{1}$.
This implies in particular that $B(2,1)$ cannot be the asymptotic symmetry group if a few generators of $\scri^{+}$
are ``missing'' or if one has ``only a piece of $\scri^{+}$''.

Therefore in order to interpret the IRS of $B(2,1)$ and relate their properties with the properties of the underlying
space$-$times we need to answer the question
which of the non$-$compact three$-$manifolds which admit flat Lorentzian metrics
also admit, when they are conformally compactified,
 future null infinity $\scri^{+}$ homeomorphic to $R \times S^{1}$. A complete answer to this question cannot be given
as these space$-$times  have not yet been studied in much detail \cite{Carlip}.


A second class of well$-$known (2+1)$-$space$-$times are those
solutions to the Einstein equations   with particle matter content.
They were first investigated in the Euclidean context by Staruszkiewicz \cite{Star} and
in
the
Lorentzian signature by Deser, Jackiw and ’t Hooft \cite{De1}.
A lot of work has been done on this class of space$-$times
since then,
e.g.,
\cite{De2},
\cite{Sou}, \cite{H1}, \cite{H2}, \cite{H3}.

The simplest  space$-$time in this class contains a single spinless point particle with mass m in $R^{3}$
at rest
at the origin. The corresponding metric was constructed in \cite{De1}.
The space is flat, but a wedge with opening angle $2 \pi (1-a)$, $a=1-4Gm$,  is cut out, and its edges are identified.
Thus the space is a cone, the unique two$-$space which is flat except at one point, its vertex. The angular defect
$2 \pi (1-a)=8 \pi G m$ characterises the mass.

The cone geometry has the following consequence: The metric admits local solutions of the killing equation corresponding
to the full Poincare group, most of the resulting would$-$be killing vectors are not preserved under the
identification of the edges of the wedge. Only the killing vectors corresponding to rotations and time translation are
globally defined.

Space$-$times with multiple point particles were
introduced in
\cite{Star}, \cite{De1},  \cite{De2}.
Curvature is created by the particles,
but only
locally at their position; elsewhere space$-$time
remains flat.
Their physical properties
have been studied
in
\cite{Star}, \cite{De1},  \cite{De2}, \cite{Sou}, \cite{H1}, \cite{H2}, \cite{H3},
for an overview see e.g. \cite{Carlip}.
Their mathematical structure is more involved than that of vacuum space$-$times
and its study was initiated in \cite{G1}, \cite{G2}, \cite{G3}.


Space$-$times with multiple point particles particles define manifolds with conical singularities or, in the case where
the masses of all particles are rational multiples of 2$\pi$, orbifolds. With the exception of the
orbifold case, they cannot be obtained as quotients of regions in three$-$dimensional Minkowski
space.

Outside point particles, space$-$time is flat. All
effects of localized sources are on the global
geometry, which is fixed by
singularities of the worldlines of the particles; these are arbitrary flat space geodesics.
This means in particular that the conserved quantities, total energy$-$momentum and
angular momentum, are related to topological invariants.
There is for example a static $N$$-$body solution \cite{De1} with
conical spatial geometry, whose total energy is additive and determines the Euler invariant of the spatial surface.

\normalfont






We await that
the IRS of $B(2,1)$ in the Hilbert topology will make contact
with both the empty space$-$times and the space$-$times with particle matter content
via the set of invariants which classify the  $SL(2,R)$$-$orbits on
 $\mathcal A^{\prime} \simeq  \mathcal H (P_{1}(R)).$


For example  the invariant which corresponds to the Poincare ``mass squa$-$ \newline red'' invariant $m^{2}$ is identified with
the mass of the particle matter content of the associated space$-$time, and subsequently,
at least in the case of the static $N$$-$body solution \cite{De1},
this invariant determines  the Euler invariant of the spatial surface  of this space$-$time.

In the case of the simplest space$-$time of a single spinless point
particle  at rest at the origin in $R^{3}$ the mass $m$ also determines the holonomy group
of the space$-$time manifold which is either a finite or a countable dense subgroup of
$SO(2)$ according to whether $1-4Gm$ is rational or not   \cite{con}.
Holonomy groups serve as  tools for the global classification of space$-$times \cite{clas} in a manner similar to but
distinct from the local Petrov and Segre type classifications.






Two$-$body quantum$-$mechanical scattering
in 2+1 gravity has been studied in
\cite{De2},  \cite{Hooft1},  \cite{Sou11}.
Since there are no gravitons forces are not
mediated by graviton exchange, rather, they are
geometrical/topological in origin, arising
from global properties of space$-$time, which is not
Minkowskian in the large, even when it is locally flat.

It is expected that the IRS of $B(2,1)$ in the nuclear topology associated with
non$-$compact little groups which leave invariant distributional supermomenta
will make contact with the
two$-$body and more generally with the N$-$body quantum$-$mechanical scattering
in 2+1 gravity. This also it is expected to happen when
$\mathcal A^{\prime}$ is enlarged to the space $\mathcal Y(S^{1})$     of real hyperfunctions on $S^{1}.$


One of the most interesting features of the representation theory of $B$ and of its generalizations in all signatures and
in all space$-$time dimensions is the appearance of discrete (finite and infinite) groups as ``little groups'' \cite{mac4}, \cite{mac6},
\cite{mac1}, \cite{Mel1}, \cite{macMel}, \cite{Mel5},  \cite{Mel7}.

In earlier papers   \cite{mac1}, \cite{McCMel}, \cite{MelProc} (and more forcibly in \cite{Mel1}) it has been argued that the IRS of $B$
and of its generalizations in complex space$-$times as well as in space$-$times with
Euclidean or Ultrahyperbolic signature induced from finite ``little groups'' are what really lie behind the full description of
(unconstrained) moduli spaces of (generalised) gravitational instantons.
It is an interesting open question to find if the IRS of $B(2,1)$ induced from
the cyclic ``little groups'' $C_{n}$ of even order
correspond to gravitational instantons in three space$-$time dimensions.

The holographic description of bulk physics in terms of a theory living on the boundary
has been concretely realised by the AdS/CFT correspondence for space$-$times with global
negative curvature. It remains an important outstanding problem to understand suitable
formulations of holography for flat space$-$time, a goal that has elicited a considerable amount
of work from several complementary approaches \cite{Bo}, \cite{Che}, \cite{Ka1}, \cite{Ka2}, \cite{St}. $B(2,1)$ is going to be relevant for
exploring holography in three$-$dimensional Minkowski space$-$time.

















\begin{thebibliography}{99}






\bibitem{Bondi}Bondi  H. \&  Van Der Berg M. G. J. \&  Metzner A. W. K.
\emph{
Gravitational waves in general relativity VII. Waves from
axi$-$symmetric isolated sources.} \normalfont
Proc. R. Soc. Lond. A.
\textbf{269},  21$-$51 (1962).














\bibitem{Sachs1}Sachs R. K.
\emph{
Gravitational waves in general relativity VIII. Waves in
asymptotically flat space$-$time.} \normalfont
 Proc. R. Soc. Lond. A.\textbf{270}, 103$-$126 (1962).


\bibitem{Pen1}Penrose R.
\emph{
Asymptotic Properties of Fields and Space$-$Times.
}. \normalfont
 Phys. Rev. Lett. \textbf{10}, 66 (1963).



\bibitem{Pen2}Penrose R.
\emph{ Relativistic symmetry groups, in: Group theory
and nonlinear problems}. \normalfont  Ed. A.O. Barut.
Reidel, Dordrecht (1976).


\bibitem{Pen3}
Penrose, R. \&  Rindler W.  \emph{  Spinors and spacetime, Vol. 2.} \normalfont
Cambridge Univ. Press, Cambridge (1986).








\bibitem{Wigner}Wigner E.
\emph{
On unitary represenations of the Inhomogeneous Lorentz
Group.} \normalfont
Annals of Mathematics \textbf{40}, 149$-$204 (1939).











\bibitem{mac3}McCarthy P. J.
\emph{
    Representations of the Bondi$-$Metzner$-$Sachs
group
\quad
I. Determination of the representations.
} \normalfont
Proc. R. Soc. Lond. A.
\textbf {330}, 517$-$535 (1972).
















\bibitem{mac8}McCarthy P. J.              \emph{
    Structure
of the  Bondi$-$Metzner$-$Sachs group.
} \normalfont
 J. Math. Phys. \textbf {13}, 1837$-$1842 (1972).



















\bibitem{mac4}McCarthy P. J.        \emph{
    Representations of the Bondi$-$Metzner$-$Sachs
group
\quad
II. Properties and classification of the  of the
representations.
}  \normalfont
 Proc. R. Soc. Lond. A. \textbf {333}, 317$-$336 (1973).








\bibitem{mac5}McCarthy P. J. \&  Crampin M.
    \emph{
    Representations of the
Bondi$-$Metzner$-$Sachs group
\quad
III. Poincare spin multiplicities and
irreducibility.
} \normalfont
Proc. R. Soc. Lond. A. \textbf {335}, 301$-$311 (1973).













\bibitem{mac6}McCarthy P. J.         \emph{
    The  Bondi$-$Metzner$-$Sachs group in
the nuclear topology.
} \normalfont
 Proc. R. Soc. Lond. A. \textbf {343}, 489$-$523 (1975).











\bibitem{mac7}McCarthy P. J. \emph{ Lifting of projective representations
of the  Bondi$-$Metzner$-$Sachs group in the nuclear topology.
} \normalfont
 Proc. R. Soc. Lond. A. \textbf {358}, 141$-$171 (1978).





















\bibitem{Crampin2}McCarthy P. J. \&
Crampin M.
\emph{
Physical significance of
the topology of the Bondi$-$Metzner$-$Sachs group.
} \normalfont
 Phys. Rev. Letters. \textbf {33}, 547$-$550 (1974).










\bibitem{Mac}McCarthy P. J.
\emph{
    Asymptotically flat Space$-$Times and Elementary Particles.
    }.  \normalfont
Phys. Rev. Lett. \textbf {29}, 817$-$819
(1972).


\bibitem{Mackey0}Mackey G. W.
\emph{
    Induced Representations of Locally Compact Groups I.
    } \normalfont
Annals of Mathematics \textbf {55}, 101$-$139
(1952).

\bibitem{Mackey00}Mackey G. W.
\emph{
    Induced Representations of Locally Compact Groups II. The
    Frobenius Reciprocity Theorem.
    } \normalfont
Annals of Mathematics \textbf {58}, 193$-$221
(1953).










\bibitem{Mackey} Mackey G. W.
\emph{
Induced representations of groups and quantum mechanics.
} \normalfont
Benjamin, New York, Amsterdam and Torino (1968).
















\bibitem{Mackey1} Mackey G. W. 1955
\emph{
The  theory of group representations.
}  \normalfont
The University of Chicago Press, Chicago (1955).













\bibitem{Simms}Simms D. J.
\emph{
Lie groups and quantum mechanics.
} \normalfont
Bonn notes
Springer, Berlin, Heidelberg and New York   (1968).













\bibitem{Isham} Isham C. J.
\emph{
Topological and quantum aspects of quantum theory.
} \normalfont
in Relativity groups and topology. North-Holland, Amsterdam (1984).





\bibitem{Mel6}  Melas E. \emph{On  the  representation  theory of  the Bondi$-$Metzner$-$Sachs group and its variants in three space$-$time dimensions}
JMP \textbf{58}, 071705 (2017).










\bibitem{Piard1}  Piard A.
\emph{
UNITARY REPRESENTATIONS OF SEMIDIRECT PRODUCT GROUPS WITH INFINITE DIMENSIONAL ABELIAN NORMAL SUBGROUP.
}  \normalfont
Rep.  Math.  Phys. \textbf{11}, 259$-$278 (1977).



\bibitem{Piard2}  Piard A.
\emph{
REPRESENTATIONS OF THE BONDI$-$METZNER$-$SACHS GROUP
WITH THE HILBERT TOPOLOGY.
}  \normalfont
Rep.  Math.  Phys. \textbf{11}, 279$-$283 (1977).












\bibitem{mac1}McCarthy P. J.
\emph{ Real and Complex Asymptotic Symmetries in Quantum Gravity, Irreducible Representations, Polygons, Polyhedra, and the A, D, E Series.}
Phil.Trans. R.  Soc. Lond. A.
\normalfont
    \textbf
{338}, 271$-$299 (1992).






















\bibitem{Mel1}  Melas E. The \emph{ BMS group and generalized gravitational instantons. } \normalfont
 J. Math. Phys. \textbf{45} \normalfont 3,  996$-$1002 (2004).




\bibitem{macMel}McCarthy P. J. \&  Melas E.
    \emph{
    On irreducible representations of the ultrahyperbolic
BMS group.
} \normalfont
Nuc. Phys. B. \textbf {653}, 369$-$399 (2003).





\bibitem{Mel5} Melas E.
\emph { Construction of the irreducibles of $B(2, 2)$.} \normalfont J.  Phys.  A:  Math.  Gen.  \textbf{39}, 3341$-$3366 (2006).


\bibitem{Melas1}  Melas E.
\emph{
Representations of the Ultrahyperbolic BMS
HB. I. General Results
}  \normalfont
(unpublished)



\bibitem{Mel6a} Melas E. \emph{ Representations of the ultrahyperbolic BMS group HB. II. Determination of the representations induced from infinite ''little groups''. }
\normalfont (2013) arXiv:1312.0532v1 (2013).

\bibitem{Mel7} Melas E. \emph{Representations of the ultrahyperbolic BMS group HB. III. Determination of the representations induced from finite ''little groups''. } \normalfont  arXiv:1402.1428v1 (2014).





\bibitem{Bi} Ashtekhar A. \&    Bi$\breve{c}$$\acute{a}$k   J. \& Schmidt  B. G.   \emph{Asymptotic structure of symmetry reduced general relativity.}   \normalfont
 Phys. Rev. D. \textbf{55} \normalfont 2,  669$-$686 (1997).






\bibitem{Ba1} Barnich  G.  \&  Troessaert   C.
\emph{Aspects of the BMS/CFT correspondence.}   \normalfont
J. High Energy Phys.   \textbf{062} \normalfont   (2010).





\bibitem{Ba2} Barnich  G.  \& Oblak  B.
\emph{Notes on the BMS group in three dimensions: I. Induced representations.}   \normalfont
J. High Energy Phys.   \textbf{129} \normalfont   (2014).





\bibitem{Ba3} Barnich  G.  \& Oblak  B.
\emph{Notes on the BMS group in three dimensions: II. Coadjoint representation.}   \normalfont
J. High Energy Phys.   \textbf{33} \normalfont   (2015).

































































































































































































\bibitem{NU}Newman E. T.   \&       Unti   W. J.
\emph{Behavior of Asymptotically Flat Empty Spaces}.
 JMP \textbf{3}, 891$-$901 (1962).





\bibitem{Sachs2}Sachs R. K.
\emph{Asymptotic Symmetries in Gravitational Theory}.
 Phys. Rev. \textbf{128}, 2851$-$2864 (1962).








\bibitem{Penrose}     Penrose R.
\emph{Spinors and space$-$time   Volume 2}. Cambridge University Press
(1986).



\bibitem{haw}Hawking  S. W. \& Perry M. J. \& Strominger A. \emph{Soft hair on black holes}.
arXiv:1601.00921v1



\bibitem{ABBOT} Giddings S. Abbot J. \& Kuchar K.
\emph{Einstein's Theory in a Three-Dimensional Space$-$Time}.
 Gen. Rel. Grav. \textbf{16}, No 8, 751$-$775 (1984).

















\bibitem{Mackey2} Mackey G. W.
\emph{Induced representations of groups and quantum mechanics}
(New york : Benjamin) (1968).



\bibitem{Mackey3}Mackey G. W.
\emph{Unitary group representations in physics, probability and
number theory. } London. Benjamin. (1978).














\bibitem{McCMel} McCarthy P. J. \& Melas E. \emph{On irreducible representations of the ultrahyperbolic BMS group}.
Nucl. \ Phys. \ B.
{\bf 653}, 369$-$399 (2003).








































\bibitem{Stone} Stone M. H. \emph{Linear transformations in Hilbert
space III, operational methods and group theory}
Proc. \ Nat. \ Acad. \ Sci. \ U.S.A. {\bf 16}, 172$–$175 (1930).








\bibitem{Cantoni}  Cantoni V. \emph{On the representations of the Bondi–Metzner–Sachs group} Ph.D. thesis, University of London (1967).




\bibitem{mac}McCarthy P. J. \& Crampin M.   \emph{Representations of the Bondi-Metzner-Sachs
group \quad   III. Poincare Spin Multiplicities and Irreducibility} Proc. R. Soc. Lond. A. \textbf {335}, 301$-$311
(1973).


\bibitem{Mackey5} Mackey G. W. \emph{Induced Representations of Groups and Quantum Mechanics} Benjamin, Amsterdam, Torino, New York,
(1968).









\bibitem{pen2}Penrose R.
\emph{Zero Rest$-$Mass Fields Including Gravitation: Asymptotic Behaviour}.
 Proc. R. Soc. Lond. A. \textbf{284}, 159$-$203 (1965).








\bibitem{Gi} Girardello L. \& Parravicini G.  \emph{Continiuous Spins in the Bondi$-$Metzner$-$Sachs
Group of Asymptotic Symmetry in General Relativity}.
Phys. \ Rev. \ Let.
{\bf 32}, 10 565$-$568 (1974).











\bibitem{Cr} Crampin M. \& McCarthy P. J.  \emph{Physical Significance of the Topology of the Bondi$-$Metzner$-$Sachs Group}.
Phys. \ Rev. \ Let.
{\bf 33}, 9 547$-$550 (1974).


\bibitem{Mc}  McCarthy P. J.  \emph{Hyperfunctions and asymptotic symmetries}.
 Proc. R. Soc. Lond. A. \textbf{358}, 495$-$498 (1978).




\bibitem{Sob} Sobolev S. L. \emph{Partial Differential Equations of Mathematical Physics}. Pergamon New York (1964).





\bibitem{Helgason}Helgason S.  \emph{Differential geometry and symmetric
spaces}.
Academic Press.    (1962).



\bibitem{Hausner}  Hausner M. \& Schwartz J. T.  \emph{Lie groups,  Lie algebras}.
Thomas Nelson and Sons Ltd London   (1968).




\bibitem{paper2} E. Melas  Representations of the
Bondi$-$Metzner$-$Sachs  group
in three space$-$time dimensions in the Hilbert topology
II. Properties and Classification of the Representations
(under preparation)



\bibitem{can} Martinec E.
\emph{Soluble systems in quantum gravity}
\normalfont
Phys. Rev. D {\bf 30},\normalfont 6,
1198$-$1204 (1984).


\bibitem{hel} Helgason  S.
\emph{Lie Groups and Symmetric Spaces}
Battelle Recontres,  p. 1,
edited by C. M.
DeWitt and J. A. Wheeler, Benjamin, New York (1968).









\bibitem{Fo} Foster J.
\emph{Conformal structure of
$\scri^{+}$
and asymptotic symmetry I. Definitions and local theory}.
J. Phys. A: Math. Gen.  {\bf 11},  1, 93$-$102 (1978).


\bibitem{Fo1} Foster J.
\emph{Asymptotic Symmetry and the Global Structure of
Future Null Infinity}.
Int. J. of Theor. Phys.  {\bf 26}, \normalfont 11, 1107$-$1124 (1987).

\bibitem{Carlip}     Carlip S.
\emph{Quantum Gravity in 2+1 Dimensions}. Cambridge University Press
(1998).

\bibitem{Star} Staruszkiewicz A.
\emph{Gravitation theory in three$-$dimensional space }.
\normalfont Acta Phys. Pol.  {\bf 24},   735$-$740 (1963).


\bibitem{De1} Deser S. \& Jackiw R. \& 't Hooft G.
\emph{Three$-$dimensional Einstein gravity: dynamics of flat space }.
\normalfont Ann. Phys.   {\bf 152},   220$-$235 (1984).


\bibitem{De2} Deser S. \& Jackiw R.
\emph{Classical and quantum scattering on a cone }.
\normalfont Commun. Math. Phys.   {\bf 118},   495$-$509 (1988).


\bibitem{Sou} de Sousa Gerbert P.
\emph{On spin and (quantum) gravity in 2+ 1 dimensions }.
\normalfont Nucl. Phys. B   {\bf 346},   440$-$472 (1990).

\bibitem{H1}  't Hooft G.
\emph{The evolution of gravitating point particles in 2+1 dimensions }.
\normalfont Class. Quantum Grav.   {\bf 10},   1023$-$1038 (1993).


\bibitem{H2}  't Hooft G.
\emph{Canonical quantization of gravitating point particles in 2+1 dimensions}.
\normalfont Class. Quantum Grav.   {\bf 10},   1653$-$1664 (1993).

\bibitem{H3}  't Hooft G.
\emph{Quantization of point particles in (2+1)-dimensional gravity and spacetime discreteness}.
\normalfont Class. Quantum Grav.   {\bf 13},   1023$-$1039 (1996).

\bibitem{G1}  Krasnov K. \& Schlenker J.
\emph{Minimal surfaces and particles in 3$-$manifolds}
\normalfont Geometriae Dedicata \bf 126 \normalfont 187$-$254
(2007)   (arXiv:math/0511441).

\bibitem{G2} Bonsante F. \& Krasnov K. \& Schlenker J.
\emph{Multi$-$black holes and earthquakes on Riemann surfaces with boundaries} Int. Math. Res. Not. \bf{2011} \normalfont  487$–$552
\normalfont
(2010)  (arXiv:math/0610429).

\bibitem{G3}  Barbot T. \& Bonsante F. \& Schlenker J.
Collisions of particles in locally AdS space$-$times: I.
Local description and global examples (2010)
(arxiv.org/abs/1010.3602v1).




\bibitem{con} Oliveira$-$Neto G.
\emph{Identifying conical singularities}
J. Math. Phys. {\bf 37} \normalfont 4716$-$4723     (1996).







\bibitem{clas} Hall G. S.  \&   Lonie D. P.
\emph{Holonomy groups and spacetimes}.
Class. Quantum Grav. {\bf 17}, \normalfont 1369$-$1382 (2000).






\bibitem{Hooft1} 't Hooft G. \emph{Non$-$Perturbative
2 Particle Scattering Amplitudes in 2+1 Dimensional Quantum
Gravity}.
Commun. Math. Phys.
{\bf 117}, \normalfont 685$-$700 (1988).

























\bibitem{Sou11} de Sousa Gerbert P. \&
Jackiw R.   \emph{Classical and Quantum Scattering on
a Spinning Cone}.
Commun. Math. Phys.
{\bf 124}, \normalfont 229$-$260 (1989).







\bibitem{MelProc}      Melas E. \emph{Approaching quantum gravity via the representation theory of the BMS
group and its variants. Some conceptual issues and a conjecture}  Nucl. Phys. B. (Proc.
Suppl.) {\bf 104}, \normalfont  212$-$216 (2002).
























\bibitem{Bo} J. de Boer  \& Solodukhin S. N. \emph{A Holographic reduction of Minkowski space$-$time}.
Nucl. \ Phys. \ B.
{\bf 665}, 545 (2003).


\bibitem{Che} Cheung C. A. de la Fuente  \& Sundrum R.  \emph{4D scattering amplitudes
and asymptotic symmetries from 2D CFT}.
JHEP
{\bf 1701}, 112 (2017).


\bibitem{Ka1} Kapec D. Mitra P. Raclariu A. M. \& Strominger A.
\emph{2D Stress Tensor for 4D Gravity}.
Phys. Rev. Lett. {\bf 119}, 12, 121601 (2017).


\bibitem{Ka2} Kapec D. Lysov V. Pasterski S. \& Strominger A.
\emph{Semiclassical Virasoro symmetry of the
quantum gravity S$-$matrix}.
JHEP {\bf 1408}, 058  (2014).

\bibitem{St} Cachazo F.   \& Strominger A.
\emph{Evidence for a New Soft
Graviton Theorem}.
arXiv: i404.4091.












































\end{thebibliography}
\end{document}